\documentclass[prc,aps,eqsecnum,twocolumn,floatfix,nofootinbib]{revtex4-1}
\usepackage{ulem}  
\usepackage{dcolumn}
\usepackage{bm}
\usepackage{color}
\usepackage{amssymb}
\usepackage{amsmath}
\usepackage{graphicx}
\usepackage{amsfonts}
\usepackage{slashed}
\usepackage{pstricks}
\usepackage{float}
\usepackage{hyperref}
\usepackage{array}
\usepackage{epsf}
\usepackage{soul}
\usepackage{siunitx}
\allowdisplaybreaks{
\begin{document}
\title{Local chiral interactions, the tritium Gamow-Teller matrix element, and the
three-nucleon contact term}
\author{A.\ Baroni$^{\rm a}$, R.\ Schiavilla$^{\rm b,c}$, L.E.\ Marcucci$^{\rm d,e}$,
L.\ Girlanda$^{\rm f,g}$, A.\ Kievsky$^{\rm e}$, A.\ Lovato$^{\rm h,i}$, S.\ Pastore$^{\,{\rm j}}$, M.\ Piarulli$^{\rm h}$, S.C.\ Pieper$^{\rm h}$, M.\ Viviani$^{\rm e}$, and R.B.\ Wiringa$^{\rm h}$}
\affiliation{
$^{\rm a}$\mbox{Department of Physics,  University of South Carolina, Columbia, South Carolina 29208, USA}\\
$^{\rm b}$\mbox{Department of Physics, Old Dominion University, Norfolk, Virginia 23529, USA}\\
$^{\rm c}$\mbox{Theory Center, Jefferson Lab, Newport News, Virginia 23606, USA}\\
$^{\rm d}$\mbox{Department of Physics, University of Pisa, 56127 Pisa, Italy}\\
$^{\rm e}$\mbox{INFN-Pisa, 56127 Pisa, Italy}\\
$^{\rm f}$\mbox{Department of Mathematics and Physics, University of Salento, 73100 Lecce, Italy} \\
$^{\rm g}$\mbox{INFN-Lecce, 73100 Lecce, Italy} \\
$^{\rm h}$\mbox{Physics Division, Argonne National Laboratory, Argonne, Illinois 60439, USA}\\
$^{\rm i}$\mbox{INFN-TIFPA, Trento Institute for Fundamental Physics and Applications, 38123 Trento, Italy}\\
$^{\rm j}$\mbox{Theoretical Division, Los Alamos National Laboratory, Los Alamos, New Mexico 87545, USA}\\
}
\date{\today}
\begin{abstract}
The Gamow-Teller (GT) matrix element contributing to tritium $\beta$ decay is calculated with
trinucleon wave functions obtained from hyperspherical-harmonics solutions of the Schr\"odinger
equation with the chiral two- and three-nucleon interactions including $\Delta$ intermediate
states that have recently been constructed in configuration space. Predictions up to
N3LO in the chiral expansion of the axial current (with $\Delta$'s) overestimate
the empirical value by 1--4 \%.  By exploiting the relation between the low-energy constant (LEC)
in the contact three-nucleon interaction and two-body axial current, we provide new determinations
of the LECs $c_D$ and $c_E$ that characterize this interaction by fitting the trinucleon binding
energy and tritium GT matrix element.  Some of the implications that the resulting models
of three-nucleon interactions have on the spectra of light nuclei and the equation of state
of neutron matter are briefly discussed.  We also provide a partial analysis, which ignores $\Delta$'s,
of the contributions due to loop corrections in the axial current at N4LO.  Finally, explicit
expressions for the axial current up to N4LO have been derived in configuration space, which
other researchers in the field may find useful.
\end{abstract}

\index{}\maketitle
\section{Introduction}
\label{sec:intro}
Tritium $\beta$ decay and the Gamow-Teller (GT) matrix element contributing
to it have provided, over the past several decades, a testing
ground for models of the nuclear axial current and, in particular, for the role
that many-body weak transition operators beyond the leading one-body GT
operator play in this matrix
element~\cite{Chemtob:1969,Riska:1970,Fishbach:1972,Carlson:1991,Schiavilla:1998}
as well as in the closely related
one entering the cross section of the basic solar burning reaction
$^1$H$(p,e^+ \nu_e)^2$H~\cite{Gari:1972,Dautry:1976,Carlson:1991,Schiavilla:1998}
{(in this connection, the first calculation of these processes in lattice
quantum chromodynamics reported last year by the NPLQCD collaboration~\cite{Savage:2017}
should also be noted).}
More recently, the development of chiral effective field theory ($\chi$EFT) has led to a
re-examination of these weak transitions within such a
framework~\cite{Park:1998,Park:2003,Marcucci:2013,Baroni:2016a,Klos:2017}
{(as well as in formulations in which the pion degrees of freedom are
integrated out---so called, pion-less effective field theory~\cite{Butler:2001,De-Leon:2016})}.
An important advantage of $\chi$EFT over older approaches based on
meson-exchange phenomenology~\cite{Chemtob:1971,Towner:1987,Riska:1989}
has been in having established a relation between the three-nucleon ($3N$) interaction
and the two-nucleon ($2N$) axial current~\cite{Gardestig:2006,Gazit:2009},
specifically between the low-energy constant (LEC) $c_D$ (in standard notation)
in the $3N$ contact interaction~\cite{Epelbaum:2002}
and the LEC in the {$2N$} contact axial current~\cite{Gazit:2009}.  Thus,
this makes it possible to use nuclear properties governed by either the strong or
weak interactions to constrain simultaneously the $3N$ interaction and $2N$
axial current.

In this context, the present study addresses two topics.  The first consists in an
assessment of how well the experimental value of the $^3$H GT matrix element is reproduced
in calculations based on nuclear Hamiltonians with the recently constructed chiral $2N$ and
$3N$ interactions~\cite{Piarulli:2016,Piarulli:2017}.  These interactions, which are local in
configuration space, have long-range parts mediated by one- and two-pion exchange (denoted as
OPE and TPE, respectively), including $\Delta$-isobar intermediate states, up to next-to-next-to-leading
order (N2LO) in the $2N$ case, and up to next-to-leading order (NLO) in the $3N$ case in the chiral
expansion.  The $2N$ and $3N$ short-range parts are parametrized by contact interactions up to,
respectively, next-to-next-to-next-to-leading order (N3LO)~\cite{Piarulli:2016} and NLO~\cite{Piarulli:2017}.
In particular, the LECs $c_D$ and $c_E$ which characterize the $3N$ contact terms have been
fitted to the trinucleon binding energies and neutron-deuteron ($nd$) doublet  scattering length.  As shown below,
the predicted GT matrix element with these interactions and accompanying axial currents
is a few \% larger than the empirical value.

The second topic deals with a determination of $c_D$ and $c_E$ in which we fit, rather than the scattering
length, the $^3$H GT matrix element.  Because of the much reduced correlation between binding energies
and the GT matrix element, this procedure leads to a more robust determination of $c_D$ and $c_E$ than
attained in the previous fit.  The axial current here includes OPE terms with $\Delta$
intermediate states up to N3LO in the chiral counting of Ref.~\cite{Baroni:2016}.  The resulting values of $c_D$ and $c_E$ are rather
different from those obtained earlier~\cite{Piarulli:2017}, and the implications that these newly calibrated
models of the $3N$ interaction have on the spectra of light nuclei and the equation of state of neutron
matter are currently being investigated~\cite{Piarulli:2018} (note that an error
in the relation given in the original Ref.~\cite{Gazit:2009} between the contact-axial-current LEC and
$c_D$ has been corrected~\cite{Schiavilla:2017}).

Related issues which we also explore in this work are (i) the magnitude of contributions
to the axial current beyond N3LO owing to loop corrections induced by TPE, and (ii) the
extent to which these contributions impact the $^3$H GT matrix
element and, in particular, modify the values of $c_D$ and $c_E$.  Since currently available derivations
of TPE axial currents in $\chi$EFT~\cite{Baroni:2016,Krebs:2017} do not explicitly
include $\Delta$'s, our comments regarding these two questions should be viewed, at this
stage, as preliminary.  Nevertheless, we believe that, even within the context of such an
incomplete analysis, it is possible to draw some conclusions, especially in reference to the
convergence pattern of the chiral expansion for the axial current.  

This paper is organized as follows.  In Sec.~\ref{sec:ax3} we list explicit expressions
in configuration space for the axial current up to N3LO.  While these are well known~\cite{Park:2003},
they are reported here for completeness and clarity of presentation, particularly in view of
the regularization scheme in configuration space that has been adopted for consistency with the
chiral interactions of Refs.~\cite{Piarulli:2016,Piarulli:2015}.  In Sec.~\ref{sec:gt3} we present
predictions for the $^3$H GT matrix element obtained with the LECs $c_D$ and $c_E$
of Ref.~\cite{Piarulli:2017}, and in Sec.~\ref{sec:fit} report a new set of values for these
LECs resulting from fitting the GT matrix element and $^3$H/$^3$He binding energies.
In Sec.~\ref{sec:ax4} we provide configuration-space expressions for the loop corrections
of the axial current at N4LO~\cite{Baroni:2016,Krebs:2017}, and estimates of their
contributions. The actual derivation of these expressions, which to the best of our knowledge
were previously not known, is relegated in Appendix~\ref{app:a1}; the resulting N4LO current
has a simple structure, which we hope will encourage its use by other researchers in the field.
Finally, we offer some concluding remarks in Sec.~\ref{sec:concl}.

\section{Axial currents up to N3LO in configuration space}
\label{sec:ax3}
\begin{figure}[bth]
 \includegraphics[width=3.25in]{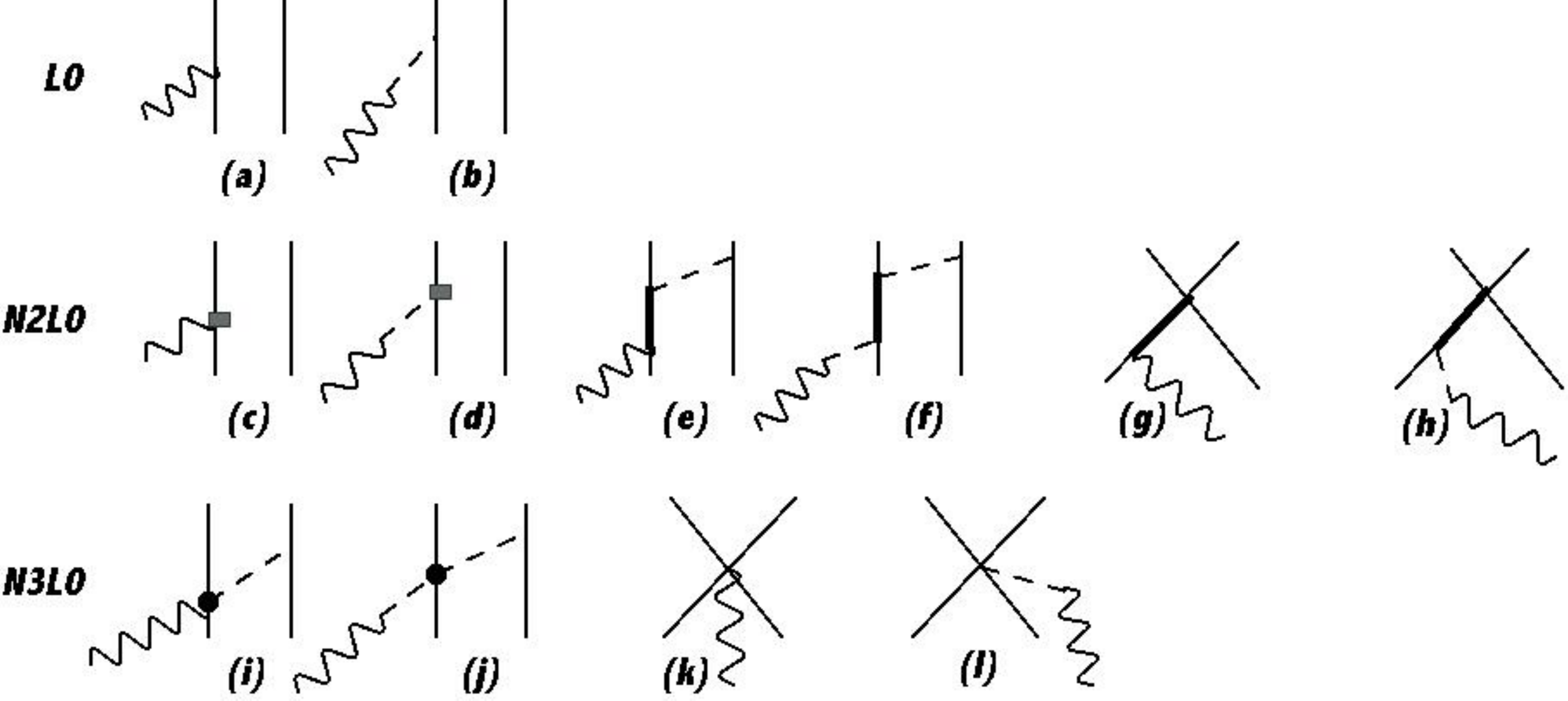}
 \caption{Diagrams illustrating the contributions to the axial current up
to N3LO ($Q^{0}$).  Nucleons, $\Delta$-isobars, pions, and external
fields are denoted by solid, thick-solid, dashed, and wavy lines,
respectively.  The squares in panel (c) and (d) represent relativistic
corrections, while the dots in panels (i) and (j) denote vertices implied by the
 ${\cal L}^{(2)}_{\pi N}$ chiral Lagrangian. Only a single time ordering is shown.
Note that the contact contributions in panels (g) and (h) vanish.}
\label{fig:f1a}
\end{figure}

We illustrate in Fig.~\ref{fig:f1a} the contributions to the axial current 
in a $\chi$EFT with nucleon, $\Delta$-isobar, and pion degrees of
freedom up to N3LO.  Momentum-space expressions for panels (a) and (b),
(c) and (d), (i) and (j), and (k) and (l) are listed in Ref.~\cite{Baroni:2016},
respectively in Eqs.~(3.14), (5.1)--(5.2), (5.5)--(5.6), and (5.4); the contributions of
panels (g) and (h) vanish, while those of panels (e) and (f) read
\begin{equation}
\label{eq:opejfin}
{\bf j}_{5,a}^{\rm N2LO}(\Delta)={\bf j}_{5,a}^{\Delta}-\frac{\bf q}{q^2+m_\pi^2}\,
{\bf q}\cdot {\bf j}_{5,a}^{\Delta} \ ,
\end{equation}
where 
\begin{eqnarray}
\label{eq:opej1fin}
{\bf j}_{5,a}^{\Delta}&=& \frac{g_A}{2\,f_\pi^2} \Big[
2\, c^\Delta_3\, \tau_{j,a}\, {\bf k}_j
+c^\Delta_4\, \left({\bm \tau}_i\times{\bm \tau}_j\right)_a 
{\bm \sigma}_i\times{\bf k}_j \Big]\nonumber\\
&&\qquad \times\, {\bm\sigma}_j\cdot{\bf k}_j\, \frac{1}{\omega_j^2}+ 
\left( i \rightleftharpoons j\right) \ ,
\end{eqnarray}
with the LECs $c_3^\Delta$ and $c_4^\Delta$ given by
\begin{equation}
c_3^\Delta=-\frac{h_A^2}{9\, m_{\Delta N}} \ , \qquad c_4^\Delta=\frac{h_A^2}{18\, m_{\Delta N}} \ .
\end{equation}
Here $g_A$ and $h_A$ are nucleon and nucleon-to-$\Delta$ axial coupling constants
($g_A\,$=$\, 1.2723$ and $h_A\,$=$\,2.74$), $f_\pi$ and $m_{\Delta N}$ are the pion-decay
constant and $\Delta$-nucleon
mass difference ($f_\pi\,$=$\, 92.4$ MeV and $m_{\Delta N}\,$=$\, 293.1$ MeV),
${\bm \sigma}_i$ and ${\bm \tau}_i$
are the spin and isospin Pauli matrices of nucleon $i$, ${\bf p}_i$ and ${\bf p}^\prime_i$ are
its initial and final momenta with the pion energy  $\omega_i$ and pion momentum  ${\bf k}_i$ defined
as $\omega_i\,$=$\, \sqrt{k_i^2+m_\pi^2}$ and ${\bf k}_i\,$=$\, {\bf p}_i^\prime-{\bf p}_i$,
and ${\bf k}_i+{\bf k}_j\,$=$\, {\bf q}$, where ${\bf q}$ is the external field momentum.

We provide below the configuration-space expressions for these
currents, ignoring pion-pole terms which
contribute negligibly to the observable under consideration in the present work.  The LO term, which
scales as $Q^{-3}$ in the power counting ($Q$ denotes generically a low-momentum scale),
reads
\begin{equation}
{\bf j}^{\rm LO}_{5,a}({\bf q})=-\frac{g_A}{2}\, \tau_{i,a}\,  {\bm \sigma}_i\, {\rm e}^{i{\bf q}\cdot {\bf r}_i} +
(i \rightleftharpoons j)\ ,
\end{equation}
while the N2LO and N3LO terms (scaling, respectively, as $Q^{-1}$  and $Q^{0}$) are written as 
\begin{eqnarray}
{\bf j}^{\rm N2LO}_{5,a}({\bf q})&=&{\bf j}^{\rm N2LO}_{5,a}({\bf q};{\rm RC})
+{\bf j}^{\rm N2LO}_{5,a}({\bf q};\Delta) \ , \\
{\bf j}^{\rm N3LO}_{5,a}({\bf q})&=& {\bf j}^{\rm N3LO}_{5,a}({\bf q};{\rm OPE})
+{\bf j}^{\rm N2LO}_{5,a}({\bf q};{\rm CT})\ ,
\end{eqnarray}
where
\begin{widetext}
\begin{eqnarray}
\label{eq:axrc}
{\bf j}^{\rm N2LO}_{5,a}({\bf q};{\rm RC})&=&\frac{g_A}{8\, m^2}\,\tau_{i,a}
\left\{{\bf p}_i\times \left( {\bm \sigma}_i \times {\bf p}_i\right)\,\, , \,\, {\rm e}^{i{\bf q}\cdot {\bf r}_i} \right\} 
+\frac{g_A}{8\, m^2}\,\tau_{i,a}\, {\rm e}^{i{\bf q}\cdot {\bf r}_i}  \left(i\,{\bf q} \times{\bf p}_i 
+{\bf q}\,\,{\bm \sigma}_i\cdot{\bf q}/2 \right) +(i\rightleftharpoons j)\ , \\
{\bf j}^{\rm N2LO}_{5,a}({\bf q};\Delta)&=&
-{\rm e}^{i\, {\bf q}\cdot {\bf r}_i}\, \left({\bm \tau}_i\times{\bm \tau}_j\right)_a \left[
I^{(1)}(\mu_{ij};\alpha_1^\Delta)\, {\bm \sigma}_i \times {\bm \sigma}_j
+I^{(2)}(\mu_{ij};\alpha_1^\Delta)\, {\bm \sigma}_i\times \hat{\bf r}_{ij}\,\, {\bm \sigma}_j\cdot \hat{\bf r}_{ij}\right] 
\nonumber\\
&&-\,{\rm e}^{i {\bf q}\cdot {\bf r}_i} \, \tau_{j,a}
\left[ I^{(1)}(\mu_{ij};\alpha_2^\Delta) \, {\bm \sigma}_j
+I^{(2)}(\mu_{ij};\alpha_2^\Delta) \,\hat{\bf r}_{ij}\,\,
 {\bm \sigma}_j\cdot \hat{\bf r}_{ij} \right]+(i\rightleftharpoons j )\ ,
\end{eqnarray}
and
\begin{eqnarray}
{\bf j}^{\rm N3LO}_{5,a}({\bf q};{\rm OPE})&=&
-{\rm e}^{i\, {\bf q}\cdot {\bf r}_i}\, \left({\bm \tau}_i\times{\bm \tau}_j\right)_a \left[
I^{(1)}(\mu_{ij};\alpha_1)\, {\bm \sigma}_i \times {\bm \sigma}_j
+I^{(2)}(\mu_{ij};\alpha_1)\, {\bm \sigma}_i\times \hat{\bf r}_{ij}\,\,{\bm \sigma}_j\cdot \hat{\bf r}_{ij}\right] 
\nonumber\\
&&-\,{\rm e}^{i {\bf q}\cdot {\bf r}_i} \, \tau_{j,a}
\left[ I^{(1)}(\mu_{ij};\alpha_2) \, {\bm \sigma}_j
+I^{(2)}(\mu_{ij};\alpha_2) \,\hat{\bf r}_{ij}\,\, {\bm \sigma}_j\cdot \hat{\bf r}_{ij} \right] \nonumber\\
&&-\left({\bm \tau}_i\times{\bm \tau}_j\right)_a\,\frac{1}{2\, m_\pi}
\left\{ {\bf p}_i \,\, , \,\, {\rm e}^{i {\bf q}\cdot {\bf r}_i}\, 
\widetilde{I}^{\,(1)}(\mu_{ij};\widetilde{\alpha}_1)\, {\bm \sigma}_j\cdot\hat{\bf r}_{ij}
 \right\} \nonumber\\
&& -i\, \left({\bm \tau}_i\times{\bm \tau}_j\right)_a\, {\rm e}^{i {\bf q}\cdot {\bf r}_i}\,
\widetilde{I}^{\,(1)}(\mu_{ij};\widetilde{\alpha}_2)\,
 {\bm \sigma}_i\times \frac{{\bf q}}{m_\pi} \,\,{\bm \sigma}_j\cdot \hat{\bf r}_{ij}
 +(i\rightleftharpoons j )\ , \\
{\bf j}^{\rm N3LO}_{5,a}({\bf q};{\rm CT})&=& z_0\, {\rm e}^{i\, {\bf q}\cdot {\bf R}_{ij}}\, \frac{ {\rm e}^{-z_{ij}^2}}{\pi^{3/2}}\,
\left({\bm \tau}_i\times{\bm \tau}_j\right)_a\,
\left({\bm \sigma}_i\times{\bm \sigma}_j\right)  \ ,
\label{eq:axct}
\end{eqnarray}
\end{widetext}
and ${\bf p}_k\,$=$\, -i\, {\bm \nabla}_k$ is the momentum operator of nucleon $k$,
$\left\{ \dots\, ,\, \dots\right\}$ denotes the anticommutator,
\begin{eqnarray}
 {\bf r}_{ij}&=&{\bf r}_i-{\bf r}_j \ ,\qquad {\bf R}_{ij}=\left({\bf r}_i+{\bf r}_j\right)/2 \ , \\
 \qquad \mu_{ij}&=& m_\pi r_{ij}\ , \qquad z_{ij}=r_{ij}/R_{\rm S} \ ,
\end{eqnarray}
and the $\delta$-function in the contact axial current has been smeared by replacing
it with a Gaussian cutoff of range $R_{\bf S}$~\cite{Piarulli:2015,Piarulli:2016}.
The adimensional LEC $z_0$ is given by
\begin{eqnarray}
\label{eq:ez0}
z_0&=&\frac{g_A}{2}\,\frac{m_\pi^2}{f_\pi^2}\, \frac{1}{\left(m_\pi \,R_{\rm S}\right)^3}
\bigg[ -\frac{m_\pi}{4\, g_A\, \Lambda_\chi}\, c_D \nonumber\\
&&+\frac{m_\pi}{3} \left( c_3+c_3^\Delta +2\, c_4+2\,c_4^\Delta \right)
+\frac{m_\pi}{6\, m} \bigg] \  ,
\end{eqnarray}
where $c_D$ denotes the LEC multiplying one of the contact terms
in the three-nucleon interaction~\cite{Epelbaum:2002}, and it should be noted
that the combination $c_3^\Delta +2\,c_4^\Delta$ vanishes. It has recently been
realized~\cite{Schiavilla:2017} that the relation between $z_0$ and $c_D$
had been given erroneously in the original reference~\cite{Gazit:2009}, a --
sign and a factor 1/4 were missing in the term proportional to $c_D$.
The various correlation functions are defined as
\begin{eqnarray}
\label{eq:e13}
I^{(1)}(\mu;\alpha)&=&-\alpha\,(1+\mu )\, \frac{e^{-\mu}}{\mu^3} \ , \\
\label{eq:e14}
I^{(2)}(\mu;\alpha)&=& \alpha\, (3+3\,\mu+\mu^2)\, \frac{e^{-\mu}}{\mu^3} \ , \\
\widetilde{I}^{\, (1)}(\mu; \widetilde{\alpha} )&=& -\widetilde{\alpha}\,(1+\mu )\, \frac{e^{-\mu}}{\mu^2} \ ,
\end{eqnarray}
where
\begin{eqnarray}
\!\!\!\!\!\!\!\!\alpha_1^\Delta&=& \frac{g_A}{8\, \pi}\, \frac{m_\pi^3}{f_\pi^2}\,c_4^\Delta\ , \,\,\,
\alpha_2^\Delta =\frac{g_A}{4\, \pi}\, \frac{m_\pi^3}{f_\pi^2}\,c_3^\Delta \ , \\
\!\!\!\!\!\!\!\!\alpha_1&=& \frac{g_A}{8\, \pi}\, \frac{m_\pi^3}{f_\pi^2}\left( c_4 +\frac{1}{4\, m}\right)\ , \,\,\,
\alpha_2 =\frac{g_A}{4\, \pi}\, \frac{m_\pi^3}{f_\pi^2}\,c_3 \ , \\
\!\!\!\!\!\!\!\!\widetilde{\alpha}_1&=& \frac{g_A}{16\, \pi}\, \frac{m_\pi^3}{m\, f_\pi^2}\ , \,\,\,
\widetilde{\alpha}_2 =\frac{g_A}{32\, \pi}\, \frac{m_\pi^3}{m\, f_\pi^2}\left(c_6+1\right)\ ,
\end{eqnarray}
$m_\pi$ and $m$ are the pion and nucleon masses, $\Lambda_\chi\,$=$\, 1$ GeV, 
and the LECs $c_3$, $c_4$, and $c_6$ have the values~\cite{Piarulli:2015,Krebs:2007}
\begin{equation}
c_3 =-0.79\,\, {\rm GeV}^{-1}   \ , \,\,c_4= 1.33\,\,\, {\rm GeV}^{-1} \ ,   \,\,\,c_6= 5.83 \ .
\end{equation}
Each correlation function above is regularized by multiplication of a configuration-space
cutoff as in the case of the local chiral potentials of Refs.~\cite{Piarulli:2015,Piarulli:2016}, namely
\begin{equation}
X^{(1,2)}(m_\pi r) \longrightarrow  C_{R_{\rm L}}(r) \, X^{(1,2)}( m_\pi r) 
\end{equation}
with
\begin{equation}
\label{eq:cfl}
C_{R_{\rm L}}(r)  = 1 - \frac{1}{ \left(r/R_{\rm L}\right)^p\, 
{\rm e}^{(r-R_{\rm L})/a_{\rm L}}+1} \ ,
\end{equation}
where $a_{\rm L}\,$=$\,R_{\rm L}/2$, the exponent $p$ is taken as $p\,$=$\, 6$, and
$X$ stands for $I$ or $\widetilde{I}$.  Finally, charge-raising ($+$) or charge-lowering ($-$)
currents are obtained from ${\bf j}_{5,\pm} = {\bf j}_{5,x}\pm i\, {\bf j}_{5,y}$, and hereafter,
we define the isospin combinations
\begin{eqnarray}
\label{eq:e22}
\tau_{i,\pm}&=&(\tau_{i,x}\pm i\, \tau_{i,y})/2 \ , \\
({\bm \tau}_1\times{\bm \tau}_2)_\pm&=&
({\bm \tau}_1\times{\bm \tau}_2)_x \pm i\, ({\bm \tau}_1\times{\bm \tau}_2)_y \ .
\end{eqnarray}

\section{$^3$H $\beta$ decay with local chiral interactions}
\label{sec:gt3}
In recent years, local chiral {$2N$} interactions have
been derived~\cite{Gezerlis:2013,Piarulli:2015,Piarulli:2016}
in configuration space, primarily for
use in quantum Monte Carlo calculations of light-nuclei and
neutron-matter properties~\cite{Gezerlis:2014,Lynn:2014,Lynn:2016,Tews:2016,Gandolfi:2017,Lynn:2017,Piarulli:2017}.
Here we focus on the
family of interactions constructed by our group~\cite{Piarulli:2016}.  These are
written as the sum of an electromagnetic-interaction component,
including up to quadratic terms in the fine-structure constant, and a strong-interaction component characterized
by long- and short-range parts.  The long-range part includes OPE and TPE terms up to N2LO in the
chiral expansion~\cite{Piarulli:2015}, derived in the static limit from leading and
sub-leading $\pi N$ and $\pi N\Delta$ chiral Lagrangians.  In coordinate
space, this long-range part is represented by charge-independent central, spin, and
tensor components with and without isospin dependence ${\bm \tau}_i\cdot {\bm \tau}_j$
(the so-called $v_6$ operator structure),
and by charge-independence-breaking central and tensor components induced by OPE
and proportional to the isotensor operator
$T_{ij}\,$=$\,3\, \tau^z_i\,\tau^z_j -{\bm \tau}_i\cdot {\bm \tau}_j$.
The radial functions multiplying these operators are singular at the origin, and
are regularized by a cutoff of the form given in Eq.~(\ref{eq:cfl}).

The short-range part is described by charge-independent contact interactions up to N3LO,
specified by a total of 20 LECs, and charge-dependent ones up to NLO, characterized
by 6 LECs~\cite{Piarulli:2016}. By utilizing Fierz identities, the resulting charge-independent
interaction can be made to contain, in addition to the $v_6$ operator structure, spin-orbit, ${\bf L}^2$ (${\bf L}$
is the relative orbital angular momentum), and quadratic spin-orbit components, while the
charge-dependent one retains central, tensor, and spin-orbit components.  Both are regularized
by multiplication of a Gaussian cutoff $C_{R_{\rm S}}(r)\,$=
$\, {\rm exp}\left[-(r/R_{\rm S})^2\right]/\left(\pi^{3/2}  R_{\rm S}^3\right)$, as in the
 contact axial current, Eq.~(\ref{eq:axct}).

Two classes of interactions were constructed, which only differ in the range of laboratory energy
over which the fits to the {$2N$} database~\cite{Perez:2013} were carried out, either 0--125 MeV
in class I or 0--200 MeV in class II.  For each class, three different sets of cutoff radii
$(R_{\rm S},R_{\rm L})$ were considered $(R_{\rm S},R_{\rm L})\,$=$\,(0.8,1.2)$ fm in set a, (0.7,1.0)
fm in set b, and (0.6,0.8) fm in set c.  The $\chi^2$/datum achieved by the fits in class I (II) was
$\lesssim 1.1$ $(\lesssim1.4)$ for a total of about 2700 (3700) data points.  We have been referring to these
high-quality {$2N$} interactions generically as the Norfolk $v_{ij}$'s (NV2s), and have been designating those
in class I as NV2-Ia, NV2-Ib, and NV2-Ic, and those in class II as NV2-IIa, NV2-IIb, and NV2-IIc.
Owing to the poor convergence of the hyperspherical-harmonics (HH) expansion
and the severe fermion-sign problem of the Green function Monte Carlo (GFMC) method, however,
models Ic and IIc have not been used (at least, not yet) in actual calculations of light nuclei.

The NV2s were found to underbind, in GFMC calculations, the ground-state energies of nuclei
with $A\,$=$\, 3$--6~\cite{Piarulli:2016}.  To remedy this shortcoming,  in Ref.~\cite{Piarulli:2017}
we constructed the leading $3N$ interaction in a $\chi$EFT, including $\Delta$ intermediate states.
It consists~\cite{Epelbaum:2002} of a long-range piece mediated by TPE at LO and NLO, and a
short-range piece parametrized in terms of two contact interactions, which enter formally at NLO.
The two (adimensional) LECs $c_D$ and $c_E$,
which characterize these latter interactions, were determined in HH calculations by simultaneously
reproducing the experimental trinucleon ground-state energies and the $nd$ doublet scattering
length for each of the {$2N$} models considered, namely, Ia and Ib, and IIa and IIb.  It was then
shown~\cite{Piarulli:2017} that the Hamiltonian based on the interactions NV2+3-Ia led, in GFMC calculations,
to an excellent description of the spectra of light nuclei in the mass range $A\,$=$\,4$--12,
including their level ordering and spin-orbit splittings.  It has since become clear~\cite{Piarulli:2018} that the
other models (NV2+3-Ib, etc.) do not provide a description of these spectra as
satisfactory as that obtained with NV2+3-Ia.

Given the value of $c_D$, the axial current is fully constrained,
since the LEC $z_0$ in the contact term is fixed via Eq.~(\ref{eq:ez0}).  The evaluation
of the tritium Gamow-Teller (GT) matrix element is carried out by Monte Carlo
integration~\cite{Baroni:2016a}, and statistical errors are less than a \% for
each individual contribution (in fact, at the level of a few parts in $10^{-4}$ for the
LO).  Predictions obtained with the Hamiltonian
models NV2+3-Ia/b and NV2+3-IIa/b and at vanishing momentum transfer
(${\bf q}\,$=$\,0$) are reported in Table~\ref{tb:tb1}. The experimental
value, as obtained in the analysis of Ref.~\cite{Baroni:2016a},
is ${\rm GT}_{\rm exp}\,$=$\,0.9511\pm 0.0013$; {it is underestimated
at LO by all models at the 3\% level, but is overestimated by $\lesssim \,$4\% in the
N3LO calculations.}
As it can be surmised from the difference between models a and b
in both classes I and II, the LO contribution is very weakly dependent
on the pair of cutoff radii ($R_{\rm S}$,$R_{\rm L}$), characterizing the
two- and three-nucleon interactions from which the $^3$H and $^3$He HH wave functions
are derived.  In contrast, the cutoff dependence is much more pronounced
in the case of the N2LO and N3LO contributions, since for these the short- and long-range
regulators  directly enter the correlation functions of the corresponding transition operators.
The N2LO(RC) correction, which is nominally
suppressed by two powers of the expansion parameter $Q/\Lambda_\chi$, being inversely
proportional to the square of the nucleon mass, itself of order $\Lambda_\chi$, is in fact further
suppressed than the naive N2LO power counting would imply.  Indeed, it is almost
an order of magnitude smaller, and of opposite sign, than the N2LO($\Delta$) contribution.
\begin{center}
\begin{table}[bth]
\begin{tabular}{l||S||S||S||S}
                         & Ia          &Ib   & IIa & IIb\\
\hline
$c_D$  & 3.666 & -2.061 & 1.278 & -4.480 \\
$c_E$  &  -1.638   &   -0.982    &   -1.029        &-0.412 \\
\hline
\hline
LO        & 0.9248  & 0.9237   &   0.9249                 & 0.9259  \\
\hline
N2LO($\Delta$)   & 0.0401     &0.0586    &   0.0406  &0.0589 \\
N2LO(RC) & -0.0055 & -0.0063 & -0.0059 &-0.0077 \\
\hline
N3LO(OPE)   & 0.0327     &  0.0457      &0.0330   & 0.0462 \\
N3LO(CT)   &  -0.0036 &   -0.0487          & -0.0249  & -0.0668\\
\hline
TOT    &       0.9885     &    0.9730        &   0.9677              &  0.9565 \\
\hline
\end{tabular}
\caption{Contributions to the 
GT matrix element in tritium $\beta$ decay obtained with chiral
axial currents up to N3LO and HH wave functions corresponding to the NV2+3-Ia/b
and NV2+3-IIa/b chiral Hamiltonians.  The experimental value
is $0.9511\pm 0.0013$~\cite{Baroni:2016a}, to be compared
to the sum of these contributions (row labeled TOT).  Also listed are the $c_D$ and $c_E$ values
of the contact terms in the three-nucleon interactions of these Hamiltonians~\cite{Piarulli:2017}.
}
\label{tb:tb1}
\end{table}
\end{center}
\vspace{-0.5cm}

The sum of the N2LO($\Delta$) and N3LO(OPE) contributions
in Table~\ref{tb:tb1} should be compared to the N3LO(OPE) contribution
reported in Ref.~\cite{Baroni:2016a} for the combinations of the Entem and Machleidt
(momentum-space) {$2N$} interactions at N3LO~\cite{Entem:2003,Machleidt:2011} and the 
Epelbaum {\it et al.}~{$3N$} interactions at LO~\cite{Epelbaum:2002} (i.e., the TPE piece
proportional to $c_1$, $c_3$, and $c_4$, and the $c_D$ and $c_E$ contact terms).
In that work, $\Delta$-isobar degrees of freedom were included implicitly, as reflected by the much
larger values (in magnitude) considered  for the LECs $c_3$ and $c_4$.  We found in Ref.~\cite{Baroni:2016a}
the N3LO(OPE) contribution to be 0.0082 (0.00043) or 0.0579 (0.0652) with the
momentum-space cutoff $\Lambda\,$=$\,500 \,\,(600)$ MeV
depending on which $c_3$-$c_4$ set was used, either the values reported
by Entem and Machleidt~\cite{Machleidt:2011} in the first case or the recent determinations by
Hoferichter and collaborators~\cite{Hoferichter:2015} in the second case.  Here, we obtain
values in the range 0.073--0.104, the lower (upper) limit corresponding to models a (b).
As we noted in Ref.~\cite{Baroni:2016a}, there are cancellations between
the individual terms proportional to $c_3$ and $c_4$, which make their sum
very sensitive to the actual values adopted for these LECs.  Nevertheless, it would
appear that the present results are close to those obtained in that work with the $c_3$ and $c_4$
values from Ref.~\cite{Hoferichter:2015}.

The magnitude (and sign) of the N3LO(CT) contribution results from the product
of the matrix element
\begin{equation}
\sum_{i\leq j}\, \langle ^3{\rm He}| \,
{\rm e}^{-z_{ij}^2}\,
\left({\bm \tau}_i\times{\bm \tau}_j\right)_+
\left({\bm \sigma}_i\times{\bm \sigma}_j\right)_z |^3{\rm H}\rangle < 0 \ ,
\end{equation}
and magnitude and sign of the LEC $z_0$, which is proportional to 
\begin{eqnarray}
z_0  &\propto& 
-\frac{m_\pi}{4\, g_A\, \Lambda_\chi}\, c_D+\frac{m_\pi}{3} \left( c_3+2\, c_4\right)
+\frac{m_\pi}{6\, m}\nonumber\\
& \simeq&  0.1105 -0.0271\, c_D \ .
\label{eq:e24}
\end{eqnarray} 
For the $c_D$ values corresponding to the interactions NV2+3-Ia/b and
NV2+3-IIa/b, we find that the N3LO(CT) contribution is negative overall.  Because
of the cancellation in $z_0$ between the constant term and the term proportional
to $c_D$ in Eq.~(\ref{eq:e24}), its magnitude is accidentally very small for model Ia.
\begin{center}
\begin{table}[bth]
\begin{tabular}{l||S||S||S||S}
                         & Ia          &Ib   & IIa & IIb\\
\hline
CT1   &  -0.0036 &   -0.0487          & -0.0249  & -0.0668\\
\hline
CT2   & -0.0037 &-0.0493 &-0.0252 &  -0.0677\\
\hline
CT3   &  -0.0036 & -0.0487     &-0.0249   & -0.0669\\
\hline
CT4   & -0.0036 & -0.0482   &-0.0246  & -0.0660 \\
\hline
\end{tabular}
\caption{
Contributions of four different parameterizations of the contact
axial current to the GT matrix element in tritium.
The first row is the same as listed in Table~\ref{tb:tb1}.
}
\label{tb:tb2}
\end{table}
\end{center}
\vspace{-0.5cm}

The N3LO(CT) contribution is only
very marginally affected by the operator structure adopted
for the contact axial current, more specifically
\begin{eqnarray}
\!\!\!\!\!\!\!{\bf j}^{\rm N3LO}_{5,+}({\rm CT1})&=&z_0\, \frac{ {\rm e}^{-z_{ij}^2}}{\pi^{3/2}}\,
\left({\bm \tau}_i\times{\bm \tau}_j\right)_+\,
\left({\bm \sigma}_i\times{\bm \sigma}_j\right) \ , \\
\!\!\!\!\!\!\!{\bf j}^{\rm N3LO}_{5,+}({\rm CT2})&=&4\, z_0\, \frac{ {\rm e}^{-z_{ij}^2}}{\pi^{3/2}}\,
\left( {\bm \sigma}_i\, \tau_{i,+} + {\bm \sigma}_j\, \tau_{j,+} \right) \ , \\
\!\!\!\!\!\!\!{\bf j}^{\rm N3LO}_{5,+}({\rm CT3})&=& 2\, z_0\, \frac{ {\rm e}^{-z_{ij}^2}}{\pi^{3/2}}\,
\left({\bm \sigma}_i - {\bm \sigma}_j\right) \left(\tau_{i,+}- \tau_{j,+} \right) \ ,\\
\!\!\!\!\!\!\!{\bf j}^{\rm N3LO}_{5,+}({\rm CT4})&=& -4\, z_0\, \frac{ {\rm e}^{-z_{ij}^2}}{\pi^{3/2}}\,
\left( {\bm \sigma}_i\, \tau_{j,+} + {\bm \sigma}_j\, \tau_{i,+} \right) \ ,
\end{eqnarray}
where the isospin-raising operators are defined as in Eq.~(\ref{eq:e22}). These structures, which
are Fierz-equivalent in the absence of the cutoff, are no longer so
when the latter is included.  The contributions corresponding to the set above
are reported in Table~\ref{tb:tb2}.
\section{Refitting $c_D$ with local chiral interactions}
\label{sec:fit}
\begin{figure}[bth]
\includegraphics[width=9cm]{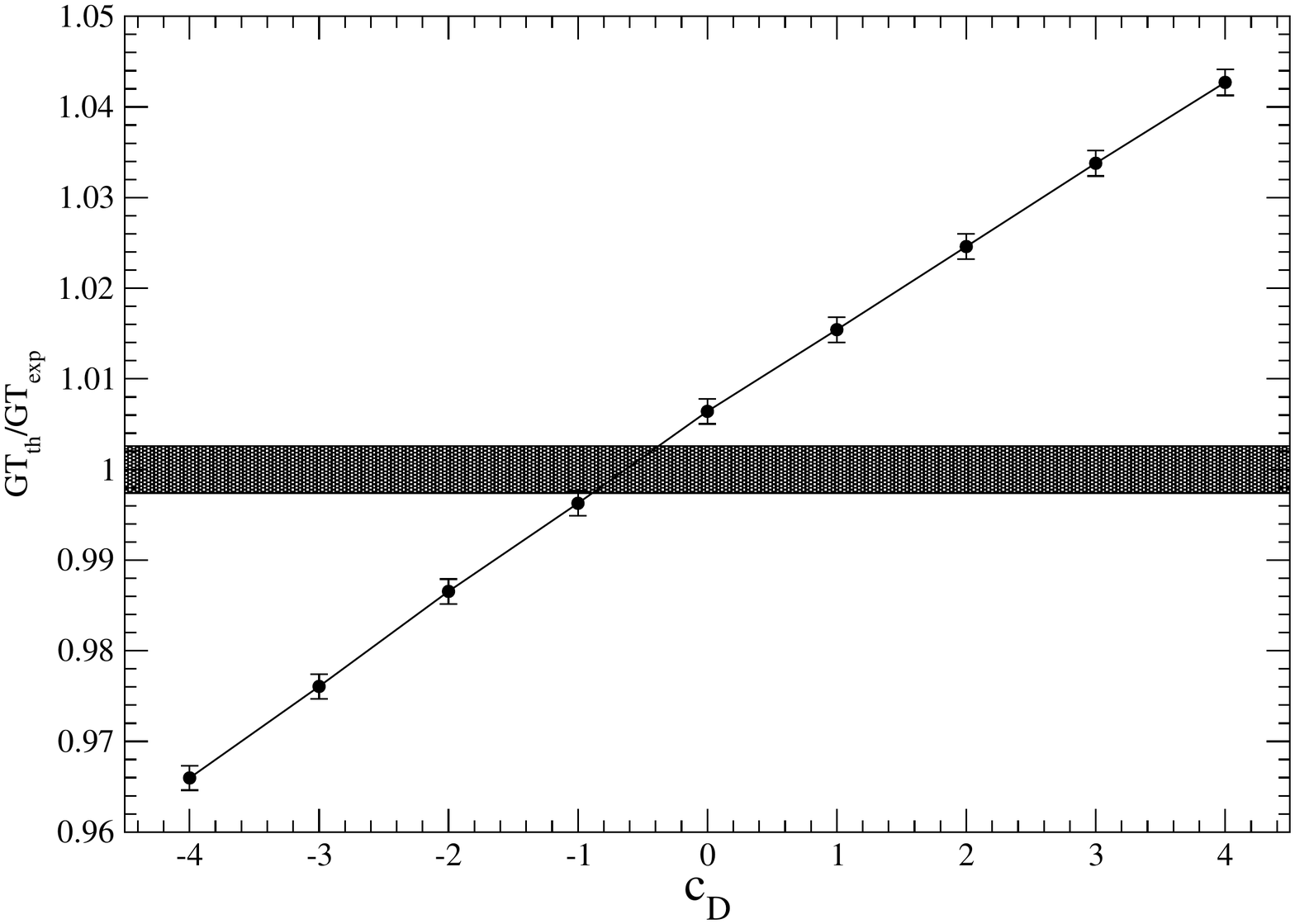}
\includegraphics[width=9cm]{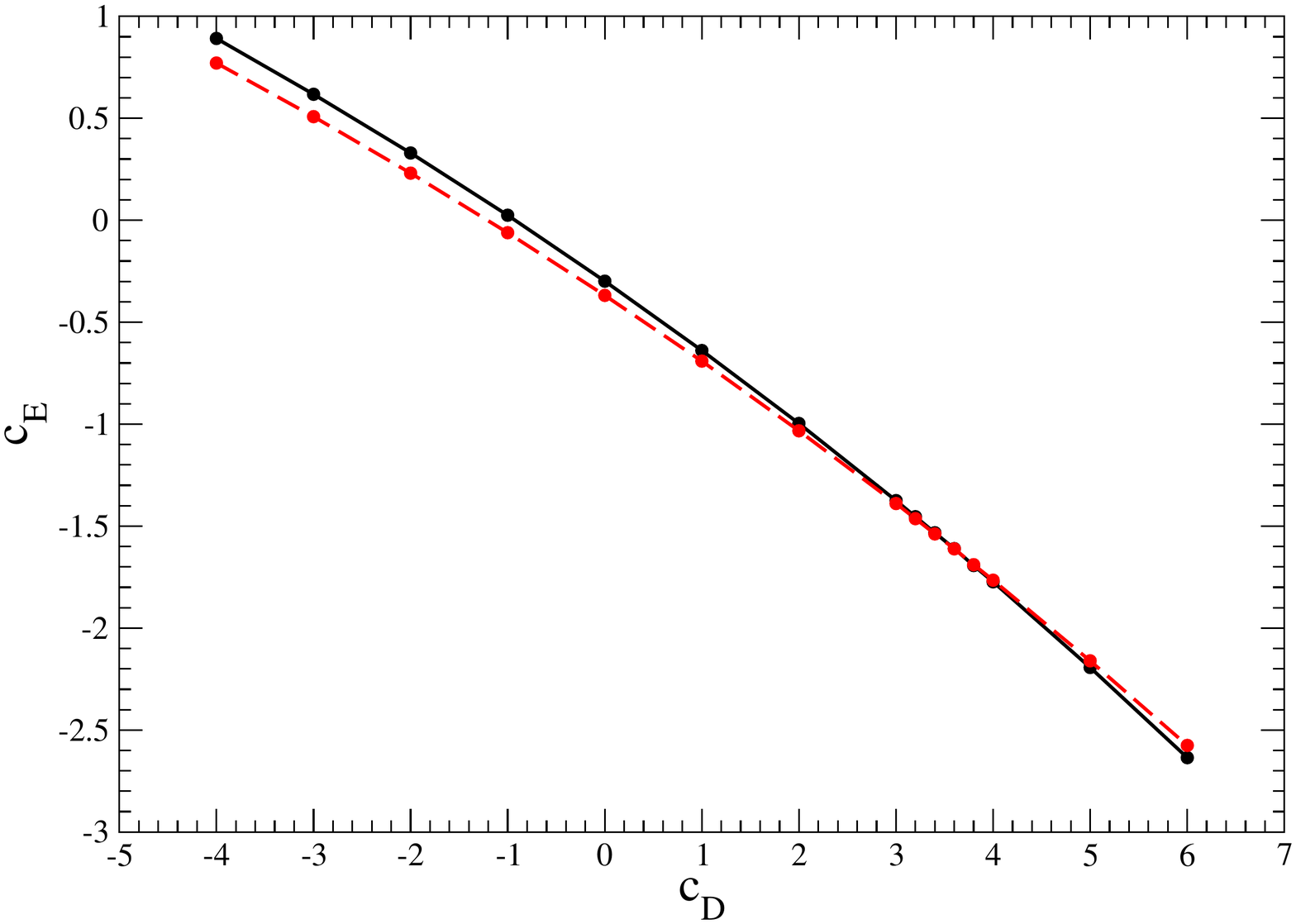}
\caption{{
Upper panel: The calculated ratio GT$_{\rm th}$/GT$_{\rm exp}$ as function of $c_D$
(solid line; each point on this line reproduces the trinucleon binding energies).
Lower panel: The $c_D$-$c_E$ trajectories obtained by fitting the experimental
trinucleon binding energies (solid line) and $nd$ doublet scattering length (dashed line)
(the intercept of these two lines gives the $c_D$ and $c_E$ values that reproduce these two
observables simultaneously). The NV2+3-Ia chiral interactions are used here for illustration.
The values of 8.475 MeV and 7.725 MeV, and $0.645\pm 0.010$ fm~\cite{Schoen:2003} 
are used for the 
$^3$H and $^3$He binding energies, and $nd$ scattering length, respectively.
Note that these energies have been corrected for the small
contributions ($+7$ keV in $^3$H and $-7$ keV in $^3$He) due to the
$n$-$p$ mass difference~\cite{Nogga:2003}. The band (left panel) results from
experimental uncertainty GT$_{\rm EXP}$, which has conservatively been doubled. 
}
}
\label{fig:ffit}
\end{figure}
In this section, we determine the LECs $c_D$ and $c_E$ in
the three-nucleon contact interaction, as parametrized in
Ref.~\cite{Piarulli:2017}, by fitting the experimental trinucleon
binding energies and central value of the $^3$H GT matrix element.
We designate these new LECs as $c_D^*$ and $c_E^*$. 
The fit is carried out as in Ref.~\cite{Baroni:2016a,Marcucci:2012}.
We span a broad range of values in $c_D$, and, in correspondence to each $c_D$ in this range, 
determine $c_E$ so as to reproduce the binding energy of either $^3$H or $^3$He. 
The resulting trajectories are nearly indistinguishable~\cite{Baroni:2016a,Marcucci:2012}.
Then, for each set of $(c_D,c_E)$, the triton and $^3$He wave functions are
calculated and the GT matrix element, denoted as GT$_{\rm th}$, is
obtained, by including in the axial current contributions up to N3LO.
The ratio GT$_{\rm th}$/GT$_{\rm exp}$ for the case of the
NV2+3-Ia interactions is shown in Fig.~\ref{fig:ffit} (left panel), where the band reflects
the uncertainty resulting from the experimental error on GT$_{\rm exp}$,
which, conservatively, has been doubled.  The LECs $(c^*_D,c^*_E)$ that
reproduce GT$_{\rm exp}$ (its central value) and the trinucleon
binding energies are reported in Table~\ref{tb:tb3}, along with the axial
current contributions at LO, N2LO, and N3LO.  In Table~\ref{tb:tb4}, we
provide the range of $(c_D^*,c_E^*)$ values compatible with the
experimental error on GT$_{\rm exp}$.  {The $3N$ interactions
corresponding to the new set of $(c_D^*,c_E^*)$ are denoted with
$^*$ hereafter.}
\begin{center}
\begin{table}[bth]
\begin{tabular}{l||S||S||S||S}
                         & Ia$^*$          &Ib$^*$   & IIa$^*$ & IIb$^*$\\
\hline
$c^*_D$  &-0.635  & -4.71 & -0.61   & -5.25 \\
\hline
$c_E^*$  &  -0.09         &  0.55         &    -0.35         & 0.05         \\
\hline
\hline
LO        & 0.9272  & 0.9247   &   0.9261                & 0.9263 \\
\hline
N2LO    & 0.0345     & 0.0517   &   0.0345  & 0.0515 \\
\hline
N3LO(OPE)   & 0.0327   &  0.0454      & 0.0330   & 0.0465\\
N3LO(CT)   &  -0.0435 &      -0.0715  & -0.0432 & -0.0737 \\
\hline
\end{tabular}
\caption{The values $c^*_D$ and $c^*_E$ obtained by fitting the
experimental trinucleon binding energies and central value of the $^3$H GT
matrix element with chiral axial currents up to N3LO and HH wave functions
corresponding to the NV2+3-Ia$^*$/b$^*$
and NV2+3-IIa$^*$/b$^*$ chiral Hamiltonians.  Also reported are the contributions
at LO, N2LO, N3LO(OPE), and N3LO(CT).
}
\label{tb:tb3}
\end{table}
\end{center}

It is interesting to compare the present $(c_D,c_E)$ trajectories (left panel of Fig.~\ref{fig:ffit})
with those of Ref.~\cite{Piarulli:2017} (right panel), obtained by fitting the experimental $nd$
doublet scattering length rather than GT$_{\rm exp}$. The strategy adopted in the present
work appears to be more robust than that of Ref.~\cite{Piarulli:2017}, since there the strong
correlation between binding energy and scattering length makes the simultaneous determination
of  $(c_D,c_E)$ problematic.  This difficulty is removed here.

The most striking difference between the previous and present determinations of LECs
is in the values of $c_E$ and $c_E^*$, in particular the fact that $c_E^*$ is quite small in
magnitude and not consistently negative as obtained in Ref.~\cite{Piarulli:2017}.  A negative
$c_E$ leads to a repulsive contribution for the associated three-nucleon contact
interaction in light nuclei~\cite{Piarulli:2017}, but to an attractive one in pure neutron matter.
Indeed, auxiliary-field diffusion Monte Carlo (AFDMC) calculations show~\cite{Piarulli:2018} that
the large and negative $c_E$ value for model NV2+3-Ia ($c_E\,$=$\,-1.638$) turns out
to be disastrous in neutron matter, since it leads to collapse already at moderate densities
(at $\rho\,$$\simeq$$\,\rho_0\,$=$\,$0.16 neutron/fm$^{3}$).
Thus, even though this model reproduces quite well the low-lying spectra of nuclei in
the mass range $A\,$=$\,$4--12, it cannot sustain the existence of neutron stars of
twice solar masses, and is therefore at variance with recent
observations~\cite{Demorest:2010,Antoniadis:2013}.  The present
determinations (the $c^*_E$'s) will mitigate, if not resolve, this issue~\cite{Piarulli:2018}.
Furthermore, because of their smallness (in magnitude), they will very substantially
reduce the cutoff dependence seen in AFDMC calculations of the neutron-matter equation of state
at high densities~\cite{Lynn:2016}.  There are also first indications that these new models,
NV2+3-Ia$^*$/b$^*$ and NV2+3-IIa$^*$/b$^*$, predict light-nuclei spectra in
{reasonable agreement} with experimental data~\cite{Piarulli:2018}.

There is a large variation between the $c_D^*$ values obtained with models NV2+3-Ia$^*$/IIa$^*$
and those with models NV2+3-Ib$^*$/IIb$^*$, which simply reflects the cutoff dependence of the
N2LO($\Delta$), N3LO(OPE), and N3LO(CT) contributions (see Table~\ref{tb:tb3}). The cutoff radii
$(R_{\rm S},R_{\rm L})$ are (0.8,1.2) fm for the former (a models) and (0.7,1.0) fm for the latter
(b models).  As a consequence, the N2LO($\Delta$) and N3LO(OPE) contributions, which both
have the same (positive) sign, increase the LO contribution and lead to an overestimate of
GT$_{\rm exp}$.  This offset is then corrected by the N3LO(CT) contribution.
In contrast to the earlier fits~\cite{Piarulli:2017}, we find the present determinations of
$c^*_D$ to be consistently negative, which make the term in the three-nucleon contact
interaction proportional to it repulsive in both light nuclei and nuclear and neutron matter.
{However, because of its one-pion leg, it is highly sensitive to tensor correlations
induced by the $2N$ interaction, so its contribution in neutron matter, where such 
correlations are weak, is noticeably reduced.}
\begin{widetext}
\begin{center}
\begin{table}[bth]
\begin{tabular}{l||c||c||c||c}
  & Ia$^*$             & Ib$^*$            & IIa$^*$                   &
  IIb$^*$\\
\hline
$c^*_D$  &  $ (-0.89,-0.38) $ & $(-4.99,-4.42)$ & $(-0.89,-0.33)$  &
$(-5.56,-4.94)$ \\
\hline
$c_E^*$  &  $(-0.01,-0.17)$ & $(+0.70,+0.40)$ & $(-0.25,-0.45)$  &
$(+0.23,-0.13)$  \\
\hline
\end{tabular}
\caption{The range of $c^*_D$ and $c^*_E$ values allowed by the experimental
error on GT$_{\rm exp}$ (note that this error has conservatively been doubled).
The lower/upper limits correspond to GT$_{\rm exp} -\!/\!+$ error.
}
\label{tb:tb4}
\end{table}
\end{center}
\end{widetext}

\section{Estimate of axial current contributions at N4LO}
\label{sec:ax4}
In this section, we provide estimates of N4LO corrections
to the GT matrix element in $^3$H.  These estimates are
incomplete, since the calculations reported here ignore
$\Delta$ intermediate states in the axial current
at N4LO.\footnote{It is useful to comment at this stage on a confusing notational
inconsistency in the power counting ascribed to interactions and currents. On the
one hand, following the customary practice in the literature, we have been referring
to two-body interaction terms of increasing order in the power counting as LO, NLO,
N2LO, and N3LO with, respectively, power scaling $Q^{0}$, $Q^{2}$, $Q^{3}$, and $Q^4$
in a two-body system, and to three-body interaction terms as LO and NLO with
scaling $Q^{-1}$ and $Q^0$ in a three-body system.
On the other hand, we denote axial-current terms as LO, N2LO, N3LO, and N4LO which scale, respectively,
as $Q^{-3}$, $Q^{-1}$, $Q^{0}$, and $Q^1$ (in a two-body system).
This notational mismatch between interactions
and currents, however, should not obscure the fact that, at least as far as the long-range
part of the interactions from OPE and TPE is concerned, there is formal consistency in the
power counting between these interactions and currents in the calculations
reported in the previous two sections.}
We are not aware of formal derivations of the two-body 
(and three-body) axial currents at this order, which include, beyond
nucleon and pion, explicit $\Delta$ degrees of freedom.
Nevertheless, it is interesting to have an approximate estimate for the
magnitude of the presently known N4LO corrections.  As a by-product of this
effort, we also obtain analytical expressions in configuration space for
these corrections, which other researchers in the field may find useful.

The (static part of the) axial current at N4LO was given in the Baroni {\it et al.}~papers
and accompanying errata~\cite{Baroni:2016a,Baroni:2016}.  It is written as the sum of
three terms: the first (labeled OPE) represents loop corrections to the OPE axial current,
the second (labeled TPE) represents genuine TPE contributions, and the last (labeled CT)
includes contact contributions induced by the regularization scheme in configuration
space we have adopted (see Appendix~\ref{app:a1} for a discussion), 
\begin{equation}
\label{eq:e29}
{\bf j}_{5,a}^{\rm N4LO}({\bf q})={\bf j}^{\rm OPE}_{5,a}({\bf q}) + {\bf j}^{\rm TPE}_{5,a}({\bf q})+{\bf j}_{5,a}^{\rm CT}({\bf q}) \ ,
\end{equation}
where 
\begin{widetext}
\begin{eqnarray}
\label{eq:e30x}
{\bf j}_{5,a}^{\rm OPE}({\bf q})&=&{\rm e}^{i {\bf q}\cdot {\bf r}_i} \, 
\frac{1}{9} \left({\bm \tau}_i\times{\bm \tau}_j\right)_a \left[
I^{(1)}(\mu_{ij};\beta)\, {\bm \sigma}_i \times {\bm \sigma}_j
+I^{(2)}(\mu_{ij};\beta)\, {\bm \sigma}_i\times \hat{\bf r}_{ij}\,\,{\bm \sigma}_j\cdot \hat{\bf r}_{ij}\right] 
\nonumber\\
&&-{\rm e}^{i {\bf q}\cdot {\bf r}_i}\, \tau_{j,a}
\left[ I^{(1)}(\mu_{ij};\beta) \, {\bm \sigma}_j
+I^{(2)}(\mu_{ij};\beta) \,\hat{\bf r}_{ij}\,\, {\bm \sigma}_j\cdot \hat{\bf r}_{ij} \right]  +(i \rightleftharpoons j)\ ,\\
\label{eq:e31x}
{\bf j}_{5,a}^{\rm TPE}({\bf q})&=&{\rm e}^{i {\bf q}\cdot {\bf r}_i} \, \tau_{j,a}
\left[ F^{(0)}_1(\lambda_{ij}) \, {\bm \sigma}_i 
-F^{(1)}_2(\lambda_{ij}) \, {\bm \sigma}_i
-F^{(2)}_2(\lambda_{ij})\,
\hat{\bf r}_{ij}\, {\bm \sigma}_i\cdot \hat{\bf r}_{ij} \right] 
-\,{\rm e}^{i {\bf q}\cdot {\bf r}_i} \, \tau_{i,a}
\left[ F^{(1)}_3(\lambda_{ij}) \, {\bm \sigma}_j
+F^{(2)}_3(\lambda_{ij}) \,\hat{\bf r}_{ij}\, {\bm \sigma}_j\cdot \hat{\bf r}_{ij} \right] \nonumber\\
\!\!\!\!\!&&\!\!\!\!\!- {\rm e}^{i\, {\bf q}\cdot {\bf R}_{ij}}\,
 \tau_{j,a}\left[ G^{(0)}_1(\lambda_{ij}) \, {\bm \sigma}_j +H_1^{(1)}(\lambda_{ij})\,  {\bm \sigma}_j
+ H^{(2)}_1(\lambda_{ij})\,
\hat{\bf r}_{ij}\, {\bm \sigma}_j\cdot \hat{\bf r}_{ij}
 \right] \nonumber\\
\!\!\!\!\!&&\!\!\!\!\!+{\rm e}^{i\, {\bf q}\cdot {\bf R}_{ij}}\!\left({\bm \tau}_i\times{\bm \tau}_j\right)_a\! \left[
H_3^{(1)}(\lambda_{ij})\, {\bm \sigma}_i \times {\bm \sigma}_j
\!+\!H_3^{(2)}(\lambda_{ij})\, {\bm \sigma}_i\times \hat{\bf r}_{ij}\left({\bm \sigma}_j\cdot \hat{\bf r}_{ij}\right)\right]\!
+\!(i\rightleftharpoons j ) , \\
\label{eq:e32x}
{\bf j}_{5,a}^{\rm CT}({\bf q}) &=& {\rm e}^{i\, {\bf q}\cdot {\bf R}_{ij}} 
\left({\bm \tau}_i\times{\bm \tau}_j\right)_a
I^{(0)}(z_{ij};\infty)\, {\bm \sigma}_i \times {\bm \sigma}_j
+\Big[ {\rm e}^{i {\bf q}\cdot {\bf r}_i}\, \tau_{j,a}\, F^{(0)}_1(z_{ij};\infty) \, 
 {\bm \sigma}_i\nonumber\\
 && - {\rm e}^{i\, {\bf q}\cdot {\bf R}_{ij}} \, \tau_{j,a} \, G^{(0)}_1(z_{ij};\infty)  \,
{\bm \sigma}_j +(i \rightleftharpoons j) \Big] \ ,
\end{eqnarray}
\end{widetext}
and pion-pole contributions are provided in Appendix~\ref{app:a1}
for completeness.  The various correlation functions, regularized by multiplication of
configuration-space cutoffs as in Sec.~\ref{sec:ax3} (and Refs.~\cite{Piarulli:2015,Piarulli:2016}),
are listed in Eqs.~(\ref{eq:e13})--(\ref{eq:e14}) and Appendix~\ref{app:a1}, Eqs.~(\ref{eq:e44})--(\ref{eq:e53})
and Eqs.~(\ref{eq:e60})--(\ref{eq:e61});
furthermore, we have defined
\begin{equation}
I^{(0)}(z_{ij};\infty)=\frac{5\, g_A^5}{1536\, \pi} \, \frac{m^4_\pi}{f_\pi^4}\,
 \frac{1}{\left(m_\pi R_{\rm S}\right)^3}\, \frac{ {\rm e}^{-z_{ij}^2}}{\pi^{3/2}} \ , 
\end{equation}
and
\begin{equation}
\beta=\frac{9\, g^5_A}{1024\, \pi^2}\, \frac{m_\pi^4}{f_\pi^4} \ , \qquad \lambda_{ij}=2\, m_\pi\, r_{ij} \ .
\end{equation}

An independent derivation of the axial current by the Bochum group in the same
$\chi$EFT framework has recently appeared in the literature~\cite{Krebs:2017}.  There
are differences at N4LO between this derivation and that of Ref.~\cite{Baroni:2016}, relating to (i) non-static two-body and static three-body contributions, which
were deliberately neglected in Ref.~\cite{Baroni:2016}, but are explicitly accounted
for in Ref.~\cite{Krebs:2017}, and (ii) a subset of static two-body contributions, specifically those obtained from box-diagram corrections
as well as loop corrections to the OPE axial current.  These differences presumably originate from the different prescriptions
adopted in these two derivations for isolating non-iterative terms in reducible diagrams. It is plausible that
the resulting forms in the two formalisms may be related to each other by a unitary
transformation~\cite{Pastore:2011}.  However, whether this is indeed the case is yet
to be established.

We report below the configuration-space expression for these differences
at vanishing momentum transfer.  We define
\begin{equation}
\Delta\,{\bf j}^{\rm N4LO}_{5,a}\equiv
{\bf j}^{\rm TOPT}_{5,a}({\bf q}\!=\!0)-{\bf j}^{\rm UT}_{5,a}({\bf q}\!=\!0)\big |_{\rm N4LO} \ ,
\end{equation}
where ${\bf j}^{\rm TOPT}_{5,a}$ and ${\bf j}^{\rm UT}_{5,a}$ are the static N4LO contributions
obtained, respectively, in Refs.~\cite{Baroni:2016a,Baroni:2016} and~\cite{Krebs:2017}, and separate
$\Delta \,{\bf j}^{\rm N4LO}_{5,a}$ as before into OPE, TPE, and associated contact terms (see Appendix~\ref{app:a1}), 
\begin{eqnarray}
\Delta \,{\bf j}^{\rm N4LO}_{5,a}&=&\Delta \,{\bf j}^{\rm N4LO}_{5,a}({\rm OPE})+\Delta \,{\bf j}^{\rm N4LO}_{5,a}({\rm TPE}) \nonumber\\
&&+\,\Delta \,{\bf j}^{\rm N4LO}_{5,a}({\rm CT}) \ ,
\end{eqnarray}
where
\begin{widetext}
\begin{eqnarray}
\label{eq:e35x}
\Delta\,{\bf j}^{\rm N4LO}_{5,a}({\rm OPE})\!&=&\!  -\frac{7}{9}\left({\bm \tau}_i\times{\bm \tau}_j\right)_a \left[
I^{(1)}(\mu_{ij};\beta)\, {\bm \sigma}_i \times {\bm \sigma}_j
+I^{(2)}(\mu_{ij};\beta)\, {\bm \sigma}_i\times \hat{\bf r}_{ij}\,\,{\bm \sigma}_j\cdot \hat{\bf r}_{ij}\right] 
\nonumber\\
&&+\frac{7}{9}\tau_{j,a}
\left[ I^{(1)}(\mu_{ij};\beta) \, {\bm \sigma}_j
+I^{(2)}(\mu_{ij};\beta) \,\hat{\bf r}_{ij}\,\, {\bm \sigma}_j\cdot \hat{\bf r}_{ij} \right]  +(i \rightleftharpoons j)\ , \\
\label{eq:e36x}
\Delta\,{\bf j}^{\rm N4LO}_{5,a}({\rm TPE})\!&=&\!- 
 \tau_{j,a}\left[ \widetilde{F}^{\,(0)}(\lambda_{ij})\, {\bm \sigma}_i
-\widetilde{G}^{\,(1)}(\lambda_{ij}) \, {\bm \sigma}_i 
-\widetilde{G}^{\,(2)}(\lambda_{ij})\,\,  \hat{\bf r}_{ij} \,\, {\bm \sigma}_i\cdot 
\hat{\bf r}_{ij} \right] + (i \rightleftharpoons j)\ , \\
\label{eq:e37x}
\Delta\,{\bf j}^{\rm N4LO}_{5,a}({\rm CT})\!&=&\!  \left({\bm \tau}_i\times{\bm \tau}_j\right)_a 
\widetilde{I}^{(0)}(z_{ij};\infty) \,
{\bm \sigma}_i \times {\bm \sigma}_j 
-\Big[ \tau_{j,a}\, \widetilde{F}^{\,(0)}(z_{ij};\infty)\, {\bm \sigma}_i
+(i \rightleftharpoons j) \Big] \ .
\label{eq:e36}
\end{eqnarray}
\end{widetext}
The correlation functions for the TPE and CT terms are listed in Appendix~\ref{app:a1},
Eqs.~(\ref{eq:e75})--(\ref{eq:e77}) and Eqs.~(\ref{eq:e79})--(\ref{eq:e80}).

The contributions of these N4LO corrections to the GT matrix element are listed
in Table~\ref{tb:tb5}.  The calculations use the HH wave functions obtained with
the Hamiltonians NV2+3Ia$^*$/b$^*$ and  NV2+3IIa$^*$/b$^*$ of the previous
section. In the table we report the ${\bf j}^{\rm N4LO}_{5,a}$ contribution as given
in Eq.~(\ref{eq:e29}) and obtained in the Baroni {\it et al.}~and Krebs {\it et al.}~formalisms,
rows labeled B and K respectively, as well as the breakup of the B contribution
into its three pieces associated with the OPE, TPE, CT terms of 
Eqs.~(\ref{eq:e30x}), (\ref{eq:e31x}), and (\ref{eq:e32x}), rows labeled OPE(B),
TPE(B), CT(B). We also provide the corresponding differences between the
B and K formalisms of Eqs.~(\ref{eq:e35x}), (\ref{eq:e36x}), and (\ref{eq:e37x}),
rows labeled B-K(OPE), B-K(TPE), and B-K(CT).
\begin{center}
\begin{table}[bth]
\begin{tabular}{l||S||S||S||S}
             & Ia$^*$          &Ib$^*$   & IIa$^*$ & IIb$^*$\\
\hline
N4LO(B)        & -0.0672 &  -0.0732  &   -0.0671                &-0.0716 \\
N4LO(K)        & -0.0364  &-0.0540    &   -0.0365         & -0.0543 \\
\hline
OPE(B)  & -0.0045    &-0.0068    & -0.0046  &-0.0069  \\
TPE(B)    &-0.0211     &-0.0326    &-0.0214   & -0.0338 \\
CT(B)  &  -0.0415   & -0.0338   & -0.0410  &-0.0310  \\
\hline  
B-K(OPE) &0.0141  &0.0196   & 0.0142  & 0.0201 \\
B-K(TPE)    &0.0018  &0.0024   & 0.0018  &0.0025  \\
B-K(CT)  & -0.0467    & -0.0412   & -0.0466  & -0.0399 \\
\hline
\end{tabular}
\caption{Contributions obtained with the Baroni {\it et al.}~\cite{Baroni:2016}  and
Krebs {\it et al.}~\cite{Krebs:2017} formulations of the N4LO axial current, denoted respectively
as N4LO(B) and N4LO(K). Also listed are the OPE(B), TPE(B), and CT(B) individual contributions
of Eqs.~(\ref{eq:e30x}), (\ref{eq:e31x}), and (\ref{eq:e32x}) in the Baroni {\it et al.} formulation,
and the corresponding differences B-K(OPE), B-K(TPE),
and B-K(CT) in the two formalisms as given in
Eqs.~(\ref{eq:e35x}), (\ref{eq:e36x}), and (\ref{eq:e37x}).
}
\label{tb:tb5}
\end{table}
\end{center}
\vspace{-0.5cm}
The contributions at N4LO are found to be relatively large and of opposite sign than
those at LO in both formalisms.  {There is
virtually no dependence on fitting the $2N$ scattering data to higher energies
(compare I to II results).}  One would expect also the N4LO contributions from the presently
ignored two-body (as well as three-body!) terms with $\Delta$ intermediate states to have a
similar magnitude and to be of the same sign {as} calculated in Table~\ref{tb:tb5}.  This makes the
convergence pattern of the chiral expansion problematic for this weak-transition
process.  It is also apparent that there is a significant cutoff dependence (compare the
$a^*$ and $b^*$ results).  Of course, this dependence could be reabsorbed into
the LEC of the contact current by enforcing agreement with the
empirical value (note that there are no additional currents of this type
that come in at N4LO).  Clearly, the values of $z_0$ (and $c_D$) would
be radically different from those listed in Table~\ref{tb:tb4}.  For example,
for the I$a^*$ case, these new $c_D$ values would be roughly 6.0 and 3.5 with, respectively, the
Baroni {\it et al.}~and Krebs {\it et al.}~estimates of the (incomplete) N4LO
corrections reported in the table above, to be compared to $c_D^*\,$=$\,-0.635$
obtained in the previous section.  Of course, these determinations
assume that $c_D$ and $c_E$ in the $3N$ contact interaction can be independently
fixed, which is only approximately valid. {Furthermore, such an analysis at N4LO would also call for the 
inclusion of loop contributions at N2LO in the $3N$ interaction.}  Finally, we have evaluated the
contribution due to one out of the many three-body axial-current mechanisms---specifically, the
expected leading term associated with TPE, panel (a) of Fig.~3 in Ref.~\cite{Baroni:2016a}---and
found it to be negligible, having values in the range \hbox{--0.0009} for Ia$^*$/IIa$^*$ to
\hbox{--0.0014} Ib$^*$/IIb$^*$.
\section{Conclusions}
\label{sec:concl}
One of questions we have examined in this work deals with the determination of the
LECs $c_D$ and $c_E$ that characterize the $3N$ interaction and nuclear
axial current, {in the context} of the chiral {$2N$ and $3N$} interaction
models with $\Delta$ intermediate states we have developed over the last
couple of years~\cite{Piarulli:2016,Piarulli:2017}.  We have shown that
$c_D$ and $c_E$ constrained to reproduce the trinucleon binding energies
and $nd$ doublet scattering length~\cite{Piarulli:2017} lead to a few \% overestimate
of the empirical value for the tritium GT matrix element.  In contrast,
the values for these LECs obtained by replacing the scattering length
with the GT matrix element in the fitting procedure (and denoted as $c^*_D$
and $c^*_E$) are very different from---and generally much smaller in
magnitude than---those above~\cite{Piarulli:2017}.
The implications of these new determinations on the spectra of
light nuclei and the equation of state of neutron matter have yet to
be fully analyzed.  However, the first indications are~\cite{Piarulli:2018} that the new
chiral Hamiltonian models NV2+3-Ia$^*$/b$^*$ and NV2+3-IIa$^*$/b$^*$ 
(with the $c_D^*$ and $c_E^*$ values in the $3N$ contact interaction)
provide a description, at least for the set of levels in the mass range $A\,$=$\,4$--10
examined so far, in reasonable accord with the observed spectra.
More importantly, the problem of neutron-matter collapse at relatively low
density, which affects, in particular, model NV2+3-Ia studied in detail in
Ref.~\cite{Piarulli:2017}, does not occur for the current models~\cite{Piarulli:2018},
since the $|c_E^*|$'s are significantly smaller than the $|c_E|$'s and, indeed, positive in some cases,
thus leading to repulsion in neutron matter for the associated (central) term
in the $3N$ contact interaction.

The other issue we have investigated concerns the size of the contribution
associated with N4LO terms in the axial current, specifically those originating from loop
corrections. Even after making allowance for current uncertainties in the
form of some of these loop corrections obtained in the Baroni {\it al.}~\cite{Baroni:2016a,Baroni:2016}
and Krebs {\it et al.}~\cite{Krebs:2017} formalisms, it appears that their contribution is
relatively large when compared to that at N2LO and N3LO, which calls into question
the convergence of the chiral expansion for the axial current. As we have {already noted},
the analysis at N4LO carried out here is incomplete, since $\Delta$ degrees
of freedom have not been accounted for consistently in either interactions or currents at
that order. Nevertheless, there is no obvious reason, at least not to us, to expect that
axial-current terms originating from TPE with $\Delta$ intermediate states would give a
contribution of opposite sign relative to that obtained currently, and so conspire to make
the overall N4LO contribution small and in line with the expected power counting.  As a
matter of fact, the convergence is already problematic in going from N2LO to N3LO (see
Table~\ref{tb:tb3}).

A future application of the interactions and currents we have developed
here will focus on the study of weak transitions---$\beta$ decays and electron- and muon-capture
processes---in light nuclei with quantum Monte Carlo methods~\cite{Pastore:2018}.
In this context, it is interesting to note that {no-core shell-model calculations of these 
transitions in the $A\,$=$\,3$--10 mass range~\cite{Navratil:2018,Hagen:2018}, based
on chiral interactions and currents,} find the sign of the overall correction beyond LO to be opposite to
that obtained for the same systems by Pastore {\it et al.}~\cite{Pastore:2018}; the exception
is tritium for which both groups find the same sign as the LO contribution. So,
the authors of Ref.~\cite{Navratil:2018,Hagen:2018} obtain a quenching of the
nuclear GT matrix elements for all these light nuclei but $^3$H {(see also
Ref.~\cite{Gazit:2009a} in connection with this issue in a calculation of $^6$He $\beta$
decay)}, while those of Ref.~\cite{Pastore:2018} always an enhancement.  It is unclear
whether this discrepancy arises from the hybrid nature of the Pastore {\it et al.}~calculation,
which used phenomenological interactions, but the chiral currents derived in Refs.~\cite{Baroni:2016a,Baroni:2016}
(albeit regularized with a momentum-space cutoff).\footnote{We note that an enhancement
was also obtained in a calculation using phenomenological interactions with two-body axial currents derived from
meson-exchange mechanisms~\cite{Pastore:2018}.}  However, one would expect the sign of the correction beyond
LO to be the same in $^3$H and the other light nuclei, as indeed obtained by Pastore {\it et al.}.
This expectation is based on the fact that, say, in a charge-raising process, the
two-body weak transition operators primarily convert a $pn$ pair with total spin-isospin $ST\,$=$\,$10
($nn$ pair with $ST\,$=$\,$01) to a $pp$ pair with $ST\,$=$\,$01 ($pn$ pair with
$ST\,$=$\,$10)~\cite{Schiavilla:1998}.  These operators, at least in light systems, do
not couple $TT_z\,$=$\,10$ to $TT_z\,$=$\,11$ in a significant way, since P-waves
are small in that case.  At small internucleon separations $\lesssim 1/m_\pi$, where
these transitions operators play a role, the pair wave functions with $ST\,$=$\,$10 and 01
in different nuclei are similar in shape and differ only by a scale factor~\cite{Forest:1996}.
Thus, the sign of these contributions should be the same.

\vspace{0.5cm}
This research is supported by the U.S.~Department of
Energy, Office of Science,  Office of Nuclear Physics, under award DE-SC0010300 (A.B.)
and contracts DE-AC05-06OR23177 (R.S.) and DE-AC02-06CH11357
(A.L.,~M.P.,~S.C.P., and~R.B.W.).
The work {of~A.L.,~S.P.,~M.P.,~S.C.P., and~R.B.W.} has been further supported by 
the NUclear Computational Low-Energy Initiative (NUCLEI) SciDAC
project.  Computational resources provided by the National Energy
Research Scientific Computing Center (NERSC) {and INFN-Pisa Computer Center
are gratefully acknowledged.}

\begin{widetext}
\appendix
\section{Axial currents at N4LO in configuration space}
\label{app:a1}
In this appendix we sketch the derivation of the configuration-space expressions for (the local part of) the axial
current at N4LO.  For completeness, we include pion-pole
contributions,
\begin{eqnarray}
{\bf j}_{5,a}^{\rm N4LO}({\bf q})&=&{\bf j}^{\rm OPE}_{5,a}({\bf q})+ {\bf j}^{\rm TPE}_{5,a}({\bf q})
+{\bf j}_{5,a}^{\rm CT}({\bf q}) 
-\frac{{\bf q}}{2\, m_\pi}\,\frac{1}{q^2/(4\, m_\pi^2)+1/4}
\left[ \frac{{\bf q}}{2\, m_\pi}\cdot {\bf j}^{\rm OPE}_{5,a}({\bf q})+\rho^{\rm TPE}_{5,a}({\bf q})+  \rho^{\rm CT}_{5,a}({\bf q})\right] 
 \ ,
\end{eqnarray}
where ${\bf j}_{5,a}^{\rm OPE}({\bf q})$, ${\bf j}_{5,a}^{\rm TPE}({\bf q})$,
and ${\bf j}_{5,a}^{\rm CT}({\bf q})$ have been defined earlier, and
\begin{eqnarray}
\!\!\!\!\!\!\rho_{5,a}^{\rm TPE}({\bf q})&=&-i\,{\rm e}^{i\, {\bf q}\cdot {\bf R}_{ij}} \, \tau_{j,a}\Big[
L^{(1)}_2(\lambda_{ij})\, {\bm \sigma}_j \cdot \hat{\bf r}_{ij} 
+L^{(1)}_1(\lambda_{ij})\left( 2\, {\bm \sigma}_i \cdot \hat{\bf r}_{ij}-{\bm \sigma}_j \cdot \hat{\bf r}_{ij}\right)
 \Big] \nonumber\\
\!\!\!\!\!\!&& +{\rm e}^{i {\bf q}\cdot {\bf r}_i} \, \tau_{j,a}\, \frac{\bf q}{2\, m_\pi}\cdot
\left[ F^{(0)}_1(\lambda_{ij}) \, {\bm \sigma}_i 
-F^{(1)}_2(\lambda_{ij}) \, {\bm \sigma}_i
-F^{(2)}_2(\lambda_{ij})\,
\hat{\bf r}_{ij}\, {\bm \sigma}_i\cdot \hat{\bf r}_{ij} \right] \nonumber\\
\!\!\!\!\!\!&&-\,{\rm e}^{i {\bf q}\cdot {\bf r}_i} \, \tau_{i,a}\, \frac{\bf q}{2\, m_\pi}\cdot
\left[ F^{(1)}_3(\lambda_{ij}) \, {\bm \sigma}_j
+F^{(2)}_3(\lambda_{ij}) \,\hat{\bf r}_{ij}\, {\bm \sigma}_j\cdot \hat{\bf r}_{ij} \right]
+ (i\rightleftharpoons j)\ ,\\
\rho_{5,a}^{\rm CT}({\bf q}) &=& -\,i\, {\rm e}^{i\, {\bf q}\cdot {\bf R}_{ij}} \, \tau_{j,a}\Big[
L^{(1)}_2(z_{ij};\infty)\, {\bm \sigma}_j \cdot \hat{\bf r}_{ij} 
+L^{(1)}_1(z_{ij};\infty)\left( 2\, {\bm \sigma}_i \cdot \hat{\bf r}_{ij}-{\bm \sigma}_j \cdot \hat{\bf r}_{ij}\right)
 \Big]  \nonumber \\
 \!\!\!\!\!\!&&+\,   \frac{{\bf q}}{2\, m_\pi}\! \cdot\! \left[{\rm e}^{i {\bf q}\cdot {\bf r}_i}\, \tau_{j,a}\, F^{(0)}_1(z_{ij};\infty) \, 
 {\bm \sigma}_i - {\rm e}^{i\, {\bf q}\cdot {\bf R}_{ij}} \, \tau_{j,a} \, G^{(0)}_1(z_{ij};\infty)  \,
{\bm \sigma}_j\right] +(i \rightleftharpoons j) \ .
\end{eqnarray}
\subsection{Loop functions}
\label{app:s1}
We begin with the momentum-space expressions in Ref.~\cite{Baroni:2016} and accompanying
errata.  After carrying out the parametric integrations, the loop functions read:
\begin{eqnarray}
\frac{1}{2\, m_\pi}\, \overline{W}_1(x)&=&-\frac{1-5\, g_A^2}{4}\, x\, {\rm arcc}(x) 
+ \frac{1-2\, g_A^2}{4\, x}\,{\rm arcs}(x)+\frac{g_A^2}{4}\frac{1}{1+x^2} \ ,\\
2\, m_\pi\,W_2(x)&=&\frac{1-g_A^2}{4} \, \frac{1}{x} \, {\rm arcs}(x)
+\frac{g_A^2}{4} \frac{1}{1+x^2}
-\frac{1+2\, g_A^2}{4\,x^2}\left[\frac{1}{x} \, {\rm arcs} (x)-1\right] \ , \\
2\, m_\pi\, W_3(x)&=& -\frac{1}{ x} \,{\rm arcs}(x) \ , \\
\frac{1}{2\, m_\pi}\, \overline{Z}_1(x)&=&- x\, {\rm arcc}(x)
+ \frac{1}{2\, x}\,{\rm arcs}(x) \ ,\\
\frac{1}{2\, m_\pi}\, \frac{\overline{Z}_2(x)}{x^2+1/4}\bigg|_{{\bf q}=0}&=& 
\frac{3}{x} \,\,\frac{3\, x^2/4+1/8}{x^2+1/4}\,\, {\rm arcs}(x)
+\, 3\,x\left[ \frac{x^2}{x^2+1/4}\,\, {\rm arcs}(x)-\frac{\pi}{2}\right]  \ ,\\ 
 \frac{1}{2\, m_\pi}\, Z_3(x)&=&\frac{1}{4}+\frac{x^2+1}{4\,x}\, {\rm arcs}(x) \ ,
\end{eqnarray}
where we have defined the adimensional variable
\begin{equation}
x=\frac{k}{2\, m_\pi}\ ,
\end{equation}
and have introduced the shorthand
\begin{equation}
{\rm arcc}(x)={\rm arccos}\frac{x}{\sqrt{1+x^2}} \qquad {\rm and}\qquad {\rm arcs}(x)={\rm arcsin}\frac{x}{\sqrt{1+x^2}} \ .
\end{equation}
The notation $Z_2(x)|_{{\bf q}=0}$ indicates that this loop function is
evaluated in the limit of vanishing momentum transfer ${\bf q}$, while the over-lines
on $W_1(x)$, $Z_1(x)$, and $Z_2(x)/(x^2+1/4)$ indicate that we have isolated a linear polynomial in $x$
in the limit $x\rightarrow \infty$ in these loop functions, that is
\begin{equation}
\overline{W}_1(x)=W_1(x)-W_1^\infty(x)\ ,
\end{equation}
and similarly for $Z_1(x)$ and $Z_2(x)/(x^2+1/4)$, where the asymptotic
polynomials read
\begin{eqnarray}
 \frac{1}{2\, m_\pi}\, W^\infty_1(x)&=&
 \frac{1-9\, g_A^2}{4} + \pi\, \frac{1-5\, g_A^2}{8} \, x \ , \\
 \frac{1}{2\, m_\pi}\,  Z_1^\infty(x)&=&1 +\frac{\pi}{2}\, x \ , \\
\frac{1}{2\, m_\pi}\,  \frac{Z_2(x)}{x^2+1/4}\bigg|^\infty_{{\bf q}=0}&=&3
+\frac{3\,\pi}{2}\, x\ .
\end{eqnarray}
\subsection{Fourier transforms}
\label{app:s2}
\begin{figure}[b]
\includegraphics[width=8cm]{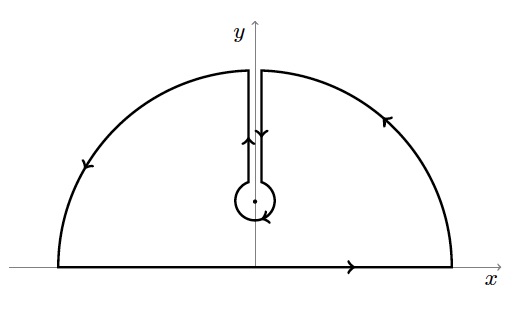}
\caption{Integration contour.}
\label{fig:fxx}
\end{figure}
In order to obtain configuration-space expressions for the N4LO
axial current, we need the following Fourier transforms of the
loop functions (with the asymptotic polynomials subtracted out
as in the previous subsection)
\begin{eqnarray}
F_i(r) =\int\frac{d{\bf k}}{(2\pi)^3}\, {\rm e}^{-i{\bf k}\cdot{\bf r}} \, W_i(k) \ ,\qquad
G_i(r) =\int\frac{d{\bf k}}{(2\pi)^3}\,  {\rm e}^{-i{\bf k}\cdot{\bf r}} \, Z_i(k) \ ,\qquad
H_i(r)=\int\frac{d{\bf k}}{(2\pi)^3}\,   {\rm e}^{-i{\bf k}\cdot{\bf r}} \,\frac{ Z_i(k)}{k^2+m_\pi^2} \ ,
\end{eqnarray}
which can be generically expressed as
\begin{equation}
\int\frac{d{\bf k}}{(2\pi)^3}\,  {\rm e}^{-i{\bf k}\cdot{\bf r}} \, f(k) = 
 \frac{(2\, m_\pi)^3}{2\, \pi^2} \,\frac{1}{\lambda} \int_0^\infty dx \, x\, {\rm sin} (x\lambda)\, f(x) \  ,
 \qquad \lambda= 2\, m_\pi\, r \ .
\end{equation}
We carry out the integrals above by utilizing contour integration in the complex plane.  We illustrate
the procedure by considering
\begin{equation}
F_3(\lambda)=\frac{(2\, m_\pi)^3}{2\, \pi^2} \,\frac{1}{\lambda} \int_0^\infty dx \, x\, {\rm sin} (x\lambda)\, W_3(x) 
=-\frac{(2\, m_\pi)^2}{2\, \pi^2}\, \frac{1}{\lambda} \, I(\lambda) \ ,
\end{equation}
where
\begin{equation}
I(\lambda)= \int_0^\infty dx\,\, {\rm sin}(x \lambda) \,\,{\rm arcsin} \frac{x}{\sqrt{1+x^2}} \ .
\end{equation}
While this integral can be done by more elementary methods,
the contour-integration technique is useful for dealing
with the more complicated transforms needed above.
By making use of the identity  ${\rm arccos}\,\alpha=\pi/2 -{\rm arcsin}\,\alpha$, we write
\begin{equation}
I(\lambda)=\frac{1}{2} \int^{\infty}_{-\infty} dx\,\, {\rm sin}(x \lambda) \,\,{\rm arcsin} \frac{x}{\sqrt{1+x^2}}
=-\frac{1}{2}\, {\rm Im}  \int^{\infty}_{-\infty} dx\,\, {\rm e}^{i\,x \lambda} \,\,{\rm arccos}
 \frac{x}{\sqrt{1+x^2}} \ ,
\end{equation}
and are then led to consider the function of the complex variable $\eta$
\begin{equation}
f(\eta)= \frac{i}{2}\,{\rm ln} \frac{\eta-i}{\eta+i}
\,\, {\rm e}^{i\,\eta \lambda_\pi}\equiv g(\eta)\,\, {\rm e}^{i\,\eta \lambda_\pi} \ ,
\end{equation}
where we have used the relation
\begin{equation}
{\rm arccos} \, \eta = -i\, {\rm ln}\left( \eta +i\, \sqrt{1-\eta^2}\right) \ .
\end{equation}
The function $f(\eta)$ has branch points at $\eta=\pm \,i$, but is otherwise analytic.  The upper cut is taken
from $i$ to $+i\,\infty$ (along the positive imaginary axis), while the lower one from
$-i$ to $-i\, \infty$ (along the negative imaginary axis). We consider the closed contour $C$ as in Fig.~\ref{fig:fxx},
so that
\begin{equation}
\oint_C d\eta\, f(\eta) =0 \ .
\label{eq:e10}
\end{equation}
Before evaluating the integral above, we need to consider the value of $f(\eta)$ to the right and left
of the cut running along the positive imaginary axis.  To this end, we define
\begin{equation}
\eta-i=r_+\, {\rm e}^{i\,\theta_+} \,\,\, {\rm with}\,\,\, -\frac{3\,\pi}{2}\leq  \theta_+ \leq \frac{\pi}{2}  \ ,
\qquad \eta+i=r_-\, {\rm e}^{i\, \theta_-} \,\,\, {\rm with}\,\,\, -\frac{\pi}{2} \leq  \theta_- \leq \frac{3\,\pi}{2} \ ,
\end{equation}
the restrictions on $\theta_\pm$ ensuring that the cuts are not crossed.  Therefore for a given $\eta$, we have
\begin{equation}
g(\eta)=\frac{i}{2}\, {\rm ln}\frac{r_+}{r_-}-\frac{\theta_+ - \theta_-}{2} \ ,
\end{equation}
and the difference along the upper cut (corresponding to $\eta=i\, y$ with $y >0 $) is given by
\begin{equation}
g(\eta)|_{{\rm left\,\,\,of\,\,\, cut}}-g(\eta)|_{{\rm right\,\,\,of\,\,\, cut}} =\pi  \ .
\end{equation}
The big arcs of radius $R$,
and the small circle of radius $r$ around the brach point $+\, i$  give 
vanishing contributions as, respectively, $R \rightarrow \infty$ and $r\rightarrow 0$.
Therefore, on the segments left and right of the {upper cut}, we find
\begin{equation}
\int_{\rm left \,\,of \,\,cut}\!\!\! \!\!\!d\eta\, f(\eta)- \int_{\rm right\,\, of\,\, cut}\!\!\! \!\!\!d\eta\, f(\eta)=
\pi \int_{i}^{\,i\, \infty} d\eta\,\,  {\rm e}^{i\,\eta \lambda} 
=i\, \pi\, \frac{{\rm e}^{-\lambda}} {\lambda} \ ,
\end{equation}
and from Eq.~(\ref{eq:e10}) we obtain
\begin{equation}
 \int^{\infty}_{-\infty} dx\,\, {\rm e}^{i\,x \lambda} \,\,{\rm arccos} \frac{x}{\sqrt{1+x^2}} + i\, \pi\, \frac{{\rm e}^{-\lambda}} {\lambda} =0 \qquad {\rm or}\qquad I(\lambda)=\frac{\pi}{2}\, \frac{{\rm e}^{-\lambda}} {\lambda} \ .
\end{equation}
By employing the integration technique above (in some instances, in addition to branch points
simple poles also occur), we find the following expressions:
 \begin{eqnarray}
\frac{1}{(2\, m_\pi)^4}\, F_1(\lambda)
&=&\frac{1}{16\,\pi} \bigg[ \left( 1-2\, g_A^2\right) \frac{1}{\lambda}
-\left(1-5\, g_A^2\right)\left(\frac{2}{\lambda^3}+\frac{2}{\lambda^2}+\frac{1}{\lambda}
-\frac{2}{\lambda^3} \, {\rm e}^{\lambda}\right)
+g_A^2 \bigg]\frac{{\rm e}^{-\lambda}}{\lambda}\ , \\
\frac{1}{(2\, m_\pi)^2}\, F_2(\lambda)&=&\frac{1}{16\,\pi} \bigg[\left( 1-g_A^2\right)\frac{{\rm e}^{-\lambda}}{\lambda^2}
 +g_A^2\,\frac{{\rm e}^{-\lambda}}{\lambda}
+\left( 1+2\,g_A^2\right)  \Gamma(-1,\lambda)\bigg] \ ,\\
\frac{1}{(2\,m_\pi)^2}\, F_3(\lambda)&=&-\frac{1}{4\,\pi}\,\,\frac{{\rm e}^{-\lambda} }{\lambda^2} \ , \\
\frac{1}{(2\, m_\pi)^4}\, G_1(\lambda)&=&-
\frac{1}{2\, \pi}  \left( \frac{1}{\lambda^2}+\frac{1}{\lambda}+\frac{1}{4} 
- \frac{{\rm e}^{\lambda}}{\lambda^2}
\right)  \frac{{\rm e}^{-\lambda}}{\lambda^2} \ ,\\
\frac{1}{(2\, m_\pi)^4}\, G_3(\lambda) &=&-\frac{1}{8\,\pi} \left(\frac{1}{\lambda}+1
- \frac{{\rm e}^{\lambda}}{\lambda}\right) 
\frac{{\rm e}^{-\lambda}}{\lambda^3}   \ ,\\
\frac{1}{(2\, m_\pi)^2}\, H_1(\lambda)&=&\frac{1}{4\,\pi} \bigg[ 
\left(1+\frac{{\rm ln}\, 3}{4} \right) \frac{{\rm e}^{-\lambda/2}}{\lambda}
 +\frac{ {\rm e}^{-\lambda}}{\lambda^2} 
 -\frac{1}{4} \int_{\lambda}^\infty dt\, 
 {\rm e}^{-t}
 \,\,\frac{1}{t^2-\lambda^2/4}\bigg]\ ,\\
 \frac{1}{(2\, m_\pi)^4} \, H_2(\lambda) &=&- \frac{9}{16\,\pi} 
\bigg[  \frac{{\rm ln}\, 3 }{12}\,\, \frac{ {\rm e}^{-\lambda/2}} {\lambda}
-\frac{{\rm e}^{-\lambda}}{\lambda^2} +\frac{1}{12} \int_{\lambda}^\infty dt\,\,{\rm e}^{-t}
\,\,  \frac{1}
{t^2-\lambda^2/4}  +\frac{4}{3\, \lambda}\, \frac{d^2}{d\lambda^2} M(\lambda)  \bigg] \ , \\
 \frac{1}{(2\, m_\pi)^2}\, H_3(\lambda)&=&\frac{1}{16\,\pi} \bigg[ 
\left(1+\frac{3\,{\rm ln}\, 3}{4} \right) \frac{{\rm e}^{-\lambda/2}}{\lambda}
 +\frac{ {\rm e}^{-\lambda}}{\lambda^2} 
 -\frac{3}{4} \int_{\lambda}^\infty dt\, 
 {\rm e}^{-t}
 \,\,\frac{1}{t^2-\lambda^2/4}\bigg]\ ,
\end{eqnarray}
where we have introduced the incomplete gamma function $\Gamma(\alpha,x)$,
 \begin{equation}
\Gamma(\alpha,x)=\int_x^\infty dt\, t^{\alpha-1}\,{\rm e}^{-t} \ ,
\end{equation}
and have defined
\begin{equation}
M(\lambda)=\frac{{\rm e}^{-\lambda }}{\lambda} -\frac{1}{\lambda} 
-\frac{{\rm ln}\, 3}{4}\, {\rm e}^{-\lambda/2} +\frac{\lambda}{4}
\int_{\lambda}^\infty dt\, {\rm e}^{-t} \, \frac{1}{t^2-\lambda^2/4} \ ,
\end{equation}
which enters $H_2(\lambda)$.  The left-over integrals and their derivatives relative to $\lambda$
can be expressed in terms of incomplete gamma functions via the identities
\begin{eqnarray}
\int_{\lambda}^\infty dt\, 
 {\rm e}^{-t}
 \,\,\frac{1}{t^2-\lambda^2/4}
 &=&\frac{{\rm e}^{-\lambda/2}}{\lambda} \, \Gamma(0,\lambda/2)
 -\frac{{\rm e}^{\lambda/2}}{\lambda} \, \Gamma(0,3\,\lambda/2) \ , \\
\frac{d}{d\lambda}\int_{\lambda}^\infty dt\, 
 {\rm e}^{-t}
 \,\,\frac{1}{t^2-\lambda^2/4}&=&-\left(1+\frac{\lambda}{2}\right)
 \frac{{\rm e}^{-\lambda/2}}{\lambda^2} \, \Gamma(0,\lambda/2)
 +\left(1-\frac{\lambda}{2}\right)
 \frac{{\rm e}^{\lambda/2}}{\lambda^2} \, \Gamma(0,3\,\lambda/2) \ ,\\
 \frac{d^2}{d\lambda^2}\int_{\lambda}^\infty dt\, 
 {\rm e}^{-t}
 \,\,\frac{1}{t^2-\lambda^2/4}&=&\left(2+\lambda+\frac{\lambda^2}{4}\right)
 \frac{{\rm e}^{-\lambda/2}}{\lambda^3} \, \Gamma(0,\lambda/2) 
 -\left(2-\lambda+\frac{\lambda^2}{4}\right)
 \frac{{\rm e}^{\lambda/2}}{\lambda^3} \, \Gamma(0,3\,\lambda/2) +\frac{{\rm e}^{\lambda}}{\lambda^2} \ .
\end{eqnarray}
\subsection{Correlation functions}
\label{app:s3}
From the Fourier transforms above, the correlation functions entering the N4LO axial current 
listed in Sec.~\ref{sec:ax4} are obtained as
\begin{eqnarray}
\label{eq:e44}
F^{(0)}_1(\lambda)&=&\frac{g_A^3}{64\, \pi}\, \frac{1}{f_\pi^4}\, F_1(\lambda)\nonumber\\
&=&\frac{g_A^3}{1024\, \pi^2}\, \frac{(2\, m_\pi)^4}{f_\pi^4}\big[  \left( 1-2\, g_A^2\right)\lambda^2_\pi
-\left(1-5\, g_A^2\right)\left(2+2\,\lambda +\lambda^2-2\,{\rm e}^{\lambda}\right)
+\,g_A^2\, \lambda^3 \big]\,\,\frac{{\rm e}^{-\lambda}}{\lambda^4}\ ,\\
\label{eq:e45}
F^{(1)}_2(\lambda) &=&\frac{g_A^3}{64\, \pi}\, 
\frac{(2\, m_\pi)^2}{f_\pi^4}\, \frac{1}{\lambda} \frac{d}{d\lambda} F_2(\lambda) \nonumber\\
&=&-\frac{g_A^3}{1024\, \pi^2}\, \frac{(2\, m_\pi)^4}{f_\pi^4}
 \big[\left( 1-g_A^2\right)\left(2+\lambda\right)
+g_A^2\,\left( \lambda+\lambda^2\right) 
+\left( 1+2\,g_A^2\right) \, \lambda \big] \,\, \frac{{\rm e}^{-\lambda}}{\lambda^4}\ , \\
\label{eq:e46}
F^{(2)}_2(\lambda) &=&\frac{g_A^3}{64\, \pi}\, \frac{(2\, m_\pi)^2}{f_\pi^4}\,
\left[  \frac{d^2}{d\lambda^2} F_2(\lambda)
-\frac{1}{\lambda} \frac{d}{d\lambda} F_2(\lambda) \right] \nonumber\\
&=& \frac{g_A^3}{1024\, \pi^2}\, \frac{(2\, m_\pi)^4}{f_\pi^4}\big[ \left( 1-g_A^2\right) \left(8+5\,\lambda+\lambda^2\right) 
+g_A^2\left(3\, \lambda+3\, \lambda^2+\lambda^3\right) 
+\,\left( 1+2\,g_A^2\right) \left(3\, \lambda+\lambda^2\right)\big] \,\, \frac{{\rm e}^{-\lambda}}{\lambda^4}\ , \\
\label{eq:e47}
F^{(1)}_3(\lambda) &=&-\frac{g_A^5}{64\, \pi}\, 
\frac{(2\, m_\pi)^2}{f_\pi^4}\left[ \frac{d^2}{d\lambda^2} F_3(\lambda) +\frac{1}{\lambda} \frac{d}{d\lambda} 
F_3(\lambda)\right] 
= \frac{g_A^5}{256\, \pi^2}\, \frac{(2\, m_\pi)^4}{f_\pi^4}\left( 4+3\, \lambda+\lambda^2\right)
 \frac{{\rm e}^{-\lambda}}{\lambda^4}\ ,\\
 \label{eq:e48}
F^{(2)}_3(\lambda) &=&\frac{g_A^5}{64\, \pi}\, 
\frac{(2\, m_\pi)^2}{f_\pi^4}\left[ \frac{d^2}{d\lambda^2} F_3(\lambda) -\frac{1}{\lambda} \frac{d}{d\lambda}
F_3(\lambda)\right] 
=- \frac{g_A^5}{256\, \pi^2}\, \frac{(2\, m_\pi)^4}{f_\pi^4}\left( 8+5\, \lambda+\lambda^2\right)
 \frac{{\rm e}^{-\lambda}}{\lambda^4}\ ,\\
 \label{eq:e49}
  G^{(0)}_1(\lambda)&=&\frac{g_A^3}{64\, \pi}\,\frac{1}{f_\pi^4}\, G_1(\lambda) 
 =-\frac{g_A^3}{128\, \pi^2}\,\frac{(2\, m_\pi)^4}{f_\pi^4}
 \left( 1+\lambda +\frac{\lambda^2}{4}-{\rm e}^\lambda \right)  \frac{{\rm e}^{-\lambda}}{\lambda^4} \ , \\
 \label{eq:e50}
 H^{(1)}_1(\lambda) &=&\frac{g_A^3}{32\, \pi}\, 
\frac{(2\, m_\pi)^2}{f_\pi^4}\, \frac{1}{\lambda} \frac{d}{d\lambda} H_1(\lambda) \nonumber\\
&=&\frac{g_A^3}{128\, \pi^2}\, 
\frac{(2\, m_\pi)^4}{f_\pi^4} \bigg[-
\left(1+\frac{{\rm ln}\, 3}{4} \right)\left(\lambda+\frac{\lambda^2}{2}\right) {\rm e}^{\lambda/2}
 -\left(2+\lambda\right)
 +\frac{1}{4}\left(\lambda+\frac{\lambda^2}{2}\right) \widetilde{\Gamma}(0,\lambda/2)\nonumber\\
&& \qquad\qquad\qquad\qquad\qquad\qquad
 -\frac{1}{4}\left(\lambda-\frac{\lambda^2}{2}\right) \widetilde{\Gamma}(0,3\,\lambda/2)
\bigg]\, \frac{ {\rm e}^{-\lambda}}{\lambda^4} \ ,\\
\label{eq:e51}
 H^{(2)}_1(\lambda) &=&\frac{g_A^3}{32\, \pi}\, 
\frac{(2\, m_\pi)^2}{f_\pi^4}\left[\frac{d^2}{d\lambda^2} H_1(\lambda) -
 \frac{1}{\lambda} \frac{d}{d\lambda} H_1(\lambda)\right] \nonumber\\
 &=&\frac{g_A^3}{128\, \pi^2}\, 
\frac{(2\, m_\pi)^4}{f_\pi^4} \bigg[
\left(1+\frac{{\rm ln}\, 3}{4} \right)\left(3\,\lambda+\frac{3\, \lambda^2}{2}
+\frac{\lambda^3}{4}\right) {\rm e}^{\lambda/2}
 +\left(8+5\,\lambda+\frac{3\, \lambda^2}{4}\right) \nonumber\\
 &&-\frac{1}{4} \left(3\,\lambda+\frac{3\, \lambda^2}{2}
+\frac{\lambda^3}{4}\right) \widetilde{\Gamma}(0,\lambda/2)
  +\frac{1}{4}\left(3\,\lambda-\frac{3\, \lambda^2}{2}
+\frac{\lambda^3}{4}\right)   \widetilde{\Gamma}(0,3\,\lambda/2)
\bigg]\, \frac{ {\rm e}^{-\lambda}}{\lambda^4} \ ,\\
\label{eq:e52}
H^{(1)}_3(\lambda) &=&\frac{g_A^3}{32\, \pi}\, 
\frac{(2\, m_\pi)^2}{f_\pi^4}\, \frac{1}{\lambda} \frac{d}{d\lambda} H_3(\lambda) \nonumber\\
&=&\frac{g_A^3}{512\, \pi^2}\, 
\frac{(2\, m_\pi)^4}{f_\pi^4} \bigg[-
\left(1+\frac{3\,{\rm ln}\, 3}{4} \right)\left(\lambda+\frac{\lambda^2}{2}\right) {\rm e}^{\lambda/2}
 -\left(2+\lambda\right)
 +\frac{3}{4}\left(\lambda+\frac{\lambda^2}{2}\right) \widetilde{\Gamma}(0,\lambda/2)\nonumber\\
&& \qquad\qquad\qquad\qquad\qquad\qquad
 -\frac{3}{4}\left(\lambda-\frac{\lambda^2}{2}\right) \widetilde{\Gamma}(0,3\,\lambda/2)
\bigg]\, \frac{ {\rm e}^{-\lambda}}{\lambda^4} \ ,\\
\label{eq:e53}
 H^{(2)}_3(\lambda) &=&\frac{g_A^3}{32\, \pi}\, 
\frac{(2\, m_\pi)^2}{f_\pi^4}\left[\frac{d^2}{d\lambda^2} H_3(\lambda) -
 \frac{1}{\lambda} \frac{d}{d\lambda} H_3(\lambda)\right] \nonumber\\
 &=&\frac{g_A^3}{512\, \pi^2}\, 
\frac{(2\, m_\pi)^4}{f_\pi^4} \bigg[
\left(1+\frac{3\,{\rm ln}\, 3}{4} \right)\left(3\,\lambda+\frac{3\, \lambda^2}{2}
+\frac{\lambda^3}{4}\right) {\rm e}^{\lambda/2}
 +\left(8+5\,\lambda+\frac{\lambda^2}{4}\right) \nonumber\\
 &&-\frac{3}{4} \left(3\,\lambda+\frac{3\, \lambda^2}{2}
+\frac{\lambda^3}{4}\right) \widetilde{\Gamma}(0,\lambda/2)
  +\frac{3}{4}\left(3\,\lambda-\frac{3\, \lambda^2}{2}
+\frac{\lambda^3}{4}\right)   \widetilde{\Gamma}(0,3\,\lambda/2)
\bigg]\, \frac{ {\rm e}^{-\lambda}}{\lambda^4} \ ,\\
\label{eq:e54}
L^{(1)}_1(\lambda)&=&\frac{g^3_A}{128\, \pi}\, \frac{1}{f_\pi^4}\, \frac{d}{d\lambda} G_1(\lambda) 
=\frac{g^3_A}{256\, \pi^2}\, \frac{(2\, m_\pi)^4}{f_\pi^4}
\left( 4+4\, \lambda+\frac{3\,\lambda^2}{2} +\frac{\lambda^3}{4}-4\, {\rm e}^\lambda\right)  
\frac{{\rm e}^{-\lambda}}{\lambda^5} \ , \\
\label{eq:e55}
L^{(1)}_2(\lambda)&=&\frac{g^3_A}{128\, \pi}\, \frac{1}{f_\pi^4}\, \frac{d}{d\lambda} H_2(\lambda) \nonumber\\
&=&  \frac{g^3_A}{512\, \pi^2}\, \frac{(2\, m_\pi)^4}{f_\pi^4}
\bigg[24+24\,\lambda+9\,\lambda^2+\frac{3\,\lambda^3}{2}-24\, {\rm e}^{\lambda} \nonumber\\
 &&+\frac{3}{8} \, \lambda^3 \left(1+\frac{\lambda}{2} \right)\widetilde{\Gamma}(0,\lambda/2)
  -\frac{3}{8}\, \lambda^3 \left(1-\frac{\lambda}{2}\right)
 \widetilde{\Gamma}(0,3\,\lambda/2)
\bigg]\, \frac{ {\rm e}^{-\lambda}}{\lambda^5} \ ,
\end{eqnarray}
where we have defined
\begin{equation}
\widetilde{\Gamma}(\alpha, x)={\rm e}^{x} \int_x^\infty dt\, t^{\alpha-1}\,{\rm e}^{-t} \ ,
\end{equation}
and $\widetilde{\Gamma}(\alpha, x)$ is computed numerically. 
The correlation functions above are regularized via
\begin{equation}
X^{(n)}_i(2\, m_\pi r) \longrightarrow  C_{R_{\rm L}}(r) \, X^{(n)}_i(2\, m_\pi r) 
\end{equation}
where $X$ stands for $F, G, H$, and $L$.

\subsection{Contact contributions}
\label{app:s4}
Asymptotic polynomials only occur in the loop functions
$W_1(k)$, $Z_1(k)$, and $Z_2(k)/(k^2+m_\pi^2)$ (see above), and
lead to contact contributions, which we regularize with the Gaussian cutoff~\cite{Piarulli:2015,Piarulli:2016}
\begin{equation}
C_{R_{\rm S}}(k) ={\rm e}^{-R^2_{\rm S}\, k^2/4} \ .
\end{equation}
We obtain
\begin{eqnarray}
\label{eq:e60}
F^{(0)}_1(z;\infty)&=&\frac{g_A^3}{128\, \pi^3}\, \frac{m^4_\pi}{f_\pi^4}\,\frac{1}{\left(m_\pi R_{\rm S}\right)^3}
\left[ \frac{1-9\, g_A^2}{2} \, C^{(0)}(z)+ \frac{1-5\, g_A^2}{8}\,\frac{\pi}{m_\pi R_{\rm S}}\, C^{(1)}(z)  \right]  \ , \\
\label{eq:e61}
G_1^{(0)}(z;\infty)&=&\frac{g_A^3}{128\, \pi^3}\, \frac{m^4_\pi}{f_\pi^4}\,\frac{1}{\left(m_\pi R_{\rm S}\right)^3}
\left[ 2\, C^{(0)}(z)+ \frac{\pi}{2\, m_\pi R_{\rm S}}\,C^{(1)}(z)  \right] \ , \\
L^{(1)}_1(z;\infty)&=&\frac{g^3_A}{512\, \pi^3}\, \frac{m_\pi^4}{f_\pi^4}
\,\frac{1}{\left(m_\pi R_{\rm S}\right)^4} \left[ 2\, \frac{d}{dz} C^{(0)}(z)+ \frac{\pi}{2\,m_\pi R_{\rm S}}\,\frac{d}{dz} C^{(1)}(z)  \right]  \ , \\
L^{(1)}_2(z;\infty)&=& \frac{g^3_A}{512\, \pi^3}\, \frac{m_\pi^4}{f_\pi^4}
\,\frac{1}{\left(m_\pi R_{\rm S}\right)^4} \left[ 6 \, \frac{d}{dz} C^{(0)}(z)+ 
\frac{3\, \pi}{2\,m_\pi R_{\rm S}}\,\frac{d}{dz} C^{(1)}(z)  \right] \ ,
\end{eqnarray}
where
\begin{eqnarray}
C^{(0)}(z)&=& 2\, \sqrt{\pi}\, {\rm e}^{-z^2} \ , \\
\frac{d}{dz} C^{(0)}(z)&=& -4\,  \sqrt{\pi}\, z\, {\rm e}^{-z^2} \ ,\\
C^{(1)}(z)&=&\int_0^\infty dx\, x^3 \, j_0(xz) \,{\rm e}^{-x^2/4}  \ , \\
\frac{d}{dz} C^{(1)}(z)&=&- \int_0^\infty dx\, x^4 \, j_1(xz) \,{\rm e}^{-x^2/4}  \ ,
\end{eqnarray}
and $z=r/R_{\rm S}$.
\subsection{Difference between Baroni {\it al.}~and Krebs~{\it et al.}}
For clarity, we report below the momentum-space expressions for the difference
between the two derivations, denoted as TOPT~\cite{Baroni:2016} and
UT~\cite{Krebs:2017} (in the limit of vanishing momentum transfer),
\begin{eqnarray}
\Delta\,{\bf j}^{\rm N4LO}_{5,a}({\bf k};{\rm OPE})&=&
-\frac{7\, g_A^5}{256\,\pi}\frac{m_\pi}{f_\pi^4}\left[ 
\tau_{j,a}\, {\bf k} -({\bm \tau}_i\times{\bm \tau}_j)_a \, {\bm \sigma}_i \times{\bf k} \right] \frac{{\bm \sigma}_j\cdot{\bf k}}{k^2+m_\pi^2}
+(i \rightleftharpoons j)\ , \\
\Delta\,{\bf j}^{\rm N4LO}_{5,a}({\bf k};{\rm TPE})&=&
-\frac{g_A^5}{128\,\pi\,f_\pi^4}\, \tau_{j,a}\left[ \widetilde{F}(k)\, {\bm \sigma}_i+\widetilde{G}(k)\, {\bf k} \, {\bm \sigma}_i\cdot{\bf k}  \right]
+(i \rightleftharpoons j)\ ,
\end{eqnarray}
where ${\bf k}\equiv{\bf k}_j=-{\bf k}_i$, and the loop functions are given by
\begin{eqnarray}
\widetilde{F}(k)&=& m_\pi \, \frac{6\,k^2+20\, m_\pi^2}{k^2+4\, m_\pi^2}  \ ,\\
\widetilde{G}(k)&=&\frac{4\, k^2+16\, m_\pi^2}{2\, k^3}\, {\rm arctan}\!\left(\! \frac{k}{2\, m_\pi}\! \right)
 - \frac{2\, m_\pi}{k^2}\, \frac{3\, k^2+8\, m_\pi^2}{k^2+4\,m_\pi^2}\ .
\end{eqnarray}
We isolate the asymptotic constant in $\widetilde{F}(k)$ as
\begin{equation}
\overline{F}(k)=\widetilde{F}(k)-\widetilde{F}^\infty \ , \qquad \overline{F}(k)=-\frac{4\, m_\pi^3 }{k^2+4\, m_\pi^2}\ , \qquad
\widetilde{F}^\infty=6\, m_\pi \ .
\end{equation}
The Fourier transforms read
\begin{eqnarray}
\widetilde{F}(\lambda)=\int_{\bf k} {\rm e}^{-i{\bf k}\cdot{\bf r}} \, \overline{F}(k) 
 &=&- \frac{(2\, m_\pi)^4}{8\, \pi} \, \frac{{\rm e}^{-\lambda}}{\lambda} \ , \\
\widetilde{G}(\lambda)= \int_{\bf k} {\rm e}^{-i{\bf k}\cdot{\bf r}} \, \widetilde{G} (k)   &=&\frac{(2\, m_\pi)^2}{2\, \pi} \left[ {\rm e}^{-\lambda}
 \left( \frac{1}{\lambda^2} -\frac{1}{2\, \lambda}\right)-\Gamma(-1,\lambda) \right] \ ,
 \end{eqnarray}
and the corresponding correlation functions are obtained as
\begin{eqnarray}
\label{eq:e75}
\widetilde{F}^{(0)}(\lambda) &=&\frac{g_A^5}{128\,\pi\,f_\pi^4} \, \overline{F}(\lambda) 
=-\frac{g_A^5}{1024\,\pi^2}\frac{(2\, m_\pi)^4}{f_\pi^4} \, \frac{{\rm e}^{-\lambda}}{\lambda} \ ,\\
\label{eq:e76}
\widetilde{G}^{(1)}(\lambda) &=&\frac{g_A^5}{128\,\pi} \, 
\frac{(2\, m_\pi)^2}{f_\pi^4}\, \frac{1}{\lambda} \frac{d}{d\lambda} \widetilde{G}(\lambda) 
=-\frac{g_A^5}{256\,\pi^2}\frac{(2\, m_\pi)^4}{f_\pi^4} \left(2-\frac{\lambda}{2}-\frac{\lambda^2}{2}\right)
\frac{{\rm e}^{-\lambda}}{\lambda^4}\ , \\
\label{eq:e77}
\widetilde{G}^{(2)}(\lambda) &=&\frac{g_A^5}{128\,\pi} \, 
\frac{(2\, m_\pi)^2}{f_\pi^4}\
\left[  \frac{d^2}{d\lambda^2} \widetilde{G}(\lambda)
-\frac{1}{\lambda} \frac{d}{d\lambda} \widetilde{G}(\lambda) \right] 
=\frac{g_A^5}{256\,\pi^2}\frac{(2\, m_\pi)^4}{f_\pi^4} \left(8+\frac{\lambda}{2}-\frac{3\, \lambda^2}{2}
-\frac{\lambda^3}{2}\right)
\frac{{\rm e}^{-\lambda}}{\lambda^4} \ .
\end{eqnarray}
We write the contact contributions from the OPE and TPE terms above as 
\begin{equation}
\!\!\Delta\,{\bf j}^{\rm N4LO}_{5,a}({\bf k};{\rm CT})\!=\! \frac{7\, g_A^5}{512\,\pi}\, 
\frac{m_\pi} {f_\pi^4} \,
({\bm \tau}_i\times{\bm \tau}_j)_a \, {\bm \sigma}_i \times {\bm \sigma}_j
-\left[\frac{g_A^5}{128\,\pi\,f_\pi^4}\, \widetilde{F}^\infty\, \tau_{j,a}\, {\bm \sigma}_i
+(i \rightleftharpoons j) \right]\ ,
\end{equation}
and define the correlation functions in Eq.~(\ref{eq:e36}) as
\begin{eqnarray}
\label{eq:e79}
 \widetilde{I}^{(0)}(z;\infty)&=&\frac{7\, g_A^5}{1024\, \pi^3}\, \frac{m^4_\pi}{f_\pi^4}\,
 \frac{1}{\left(m_\pi R_{\rm S}\right)^3}\, C^{(0)}(z) \ ,\\
 \label{eq:e80}
 \widetilde{F}^{(0)}(z;\infty)&=&\frac{3\, g_A^5}{128\, \pi^3}\, \frac{m^4_\pi}{f_\pi^4}\,
 \frac{1}{\left(m_\pi R_{\rm S}\right)^3}\, C^{(0)}(z) \ .
\end{eqnarray}
\end{widetext}

\end{document}